\documentclass{aa}

\usepackage{graphicx}
\usepackage{txfonts}
\usepackage{xspace}
\usepackage{tablefootnote}
\usepackage{threeparttable}
\usepackage{siunitx} 
\usepackage{booktabs} 
\usepackage{hhline} 
\usepackage{amsmath} 
\usepackage{hyperref}
\usepackage{caption}
\usepackage{subcaption}

\DeclareMathOperator{\erf}{erf}

\begin{document}

   \title{In-flight calibration of the INTEGRAL/IBIS Compton mode}
   \subtitle{Application to the Crab Nebula polarization}
   \titlerunning{IBIS Compton mode polarization}
   
   \author{T. Bouchet \inst{1,2} \and
          P. Laurent \inst{1} \and
         F. Cangemi \inst{3} \and
          J. Rodriguez \inst{1}
          }

  \institute{AIM, CEA/CNRS/Universit\'e Paris-Saclay, Universit\'e Paris Cit\'e, F-91191 Gif-sur-Yvette, France
  \and Julius-Maximilians-Universität Würzburg, Fakultät für Physik und Astronomie, Institut für Theoretische Physik und Astrophysik, Lehrstuhl für Astronomie, Emil-Fischer-Str. 31, 97074 Würzburg, Germany
    \and APC, Universit\'e Paris Cit\'e/CNRS/CEA, 75013 Paris, France
             }

   \date{\today}
 
  \abstract
   {The INTEGRAL satellite explored the $\gamma$-ray sky since its launch on October 17$^{th}$, 2002, and until the end of its scientific operation on February 28$^{th}$, 2025. A large fraction of the available data is still largely untouched, due to the complexity of analysis.}
   {We describe the latest in-flight calibration of the Compton mode of the INTEGRAL/IBIS telescope, taking into account more than twenty years of data. The spectroscopy and polarization of the standard candle that is the Crab Nebula is analyzed in detail.}
   {We operate the IBIS telescope as a Coded mask Compton telescope, using the Crab Nebula to refine the calibration, as is usually done for high-energy instruments.}
   {We have determined the spectroscopic and polarimetric properties of the IBIS Compton mode and their evolution along the entire duration of the mission. In addition, the long-term evolution of the Crab Nebula's polarization has been successfully measured and compared with other high-energy experiments. We could estimate the energy dependence of the Crab Nebula polarization in four bands between 200\,keV and 1\,MeV. In particular, the detection of polarized emissions strictly above 400\,keV makes it the highest energy measurement ever performed for the Crab Nebula. A Python library was also made publicly available to analyze processed data.}
  {}
   \keywords{Gamma rays: general - Polarization - stars: individual: Crab Nebula}

\maketitle

\section{Introduction}

Polarization measurements of astrophysical sources in the soft $\gamma$-rays (100\,keV--100\,MeV) is a very recent scientific topic which only started in the early 2000s, more than 40 years after the first high-energy space telescopes were launched \citep{Giacconi_1962}. This is partly due to the sophisticated methods involved, as well as the faintness of most sources at those energies, requiring highly sensitive detectors. Despite the difficulties involved, measuring polarization is a worthwhile objective as it is a crucial tool to understand the emission processes in high-energy sources \citep{Zdziarski_2012, review_polar_jet}.\\

The first measurement was done on the Crab Nebula by the SPI instrument onboard INTEGRAL \citep{Dean_2008}, and was shortly confirmed by the IBIS telescope onboard the same satellite \citep{Forot_2008}. The same methods were than applied to other high-energy sources, namely X-ray binaries and Gamma-Ray Bursts (GRB). Other experiments involving different missions followed, further increasing the number of results \citep[see a review by][]{Chattopadhyay_2021}.\\

The only astrophysical source considered here is the Crab Nebula (RA=05$^h$34$^m$31.8$^s$, dec=\ang{+22;01;03}) -- a Supernova remnant with a young pulsar at its core. Both the pulsar and nebula are known to produce polarized synchrotron emission in optical \citep{crab_optical_polar}, X-rays \citep{ixpe_crab_polar}, and $\gamma$-rays \citep{Dean_2008, Forot_2008}. The polarization along with the spectrum can prove very useful to probe the geometry of the emission medium. In particular, the fraction of polarization traces the degree of uniformity in the magnetic fields. This, in turn, can constrain the model for pulsar emission, which is still a major problem in astronomy \citep{Moran_2013}.\\

This paper first gives an overview of the INTEGRAL/IBIS Compton mode, which measures spectra and (linear) polarization in the hard X-rays to soft $\gamma$-rays (Section \ref{sec:description}). Although the general method is already explained in \citet{Forot_2007}, we intend to give here a more detailed explanation of the software itself, including the latest changes (Section \ref{sec:software}). We explain the calibration of the count-rate estimation using the Crab Nebula (Crab hereafter) as a "standard candle", as is done in many space experiment at high-energies (Section \ref{sec:calibration}). Finally, we apply our method to the Crab on nearly the entire archive of INTEGRAL observations (Section \ref{sec:polar}), revealing new information about this astrophysical source (Section \ref{sec:discussion}). In particular, we achieved an unprecedented measurement in the 400 -- 1000\,keV range, probing further into the $\gamma$-ray polarization of the Crab Nebula. The data is also analyzed in energy and time, and compared with other independent measurements in order to show the robustness of our method.

\section{Description of INTEGRAL/IBIS Compton mode}\label{sec:description}

\subsection{Principles of Coded Mask Compton Telescopes}\label{sec:theory}

In a Compton telescope consisting of two superimposed detector layers, $\gamma$-ray photons with an initial energy $E_0$ are Compton scattered in one detector and absorbed in the other. In our case, we focus on the forward Compton scattering, where the first interaction occurs in the upper-plane (with respect to the pointing direction), and subsequent absorption in the lower-plane. The locations and energy deposits of each interaction are measured, and two events are associated if they are both detected within the same coincidence time window. The energies deposited can be deduced with the proper energy calibration, and interpreted as the recoil electron energy in the upper-plane ($E_1$), and scattered photon energy in the lower-plane ($E_2$) in the case of a forward Compton event. The positions of the interactions inform us about the polar diffusion angle ($\theta$) and the azimuthal diffusion angle with respect to the North-East direction ($\phi$). Fig. \ref{fig:sketch_ibis_compton} shows the geometry of a Compton interaction in the case of an on-axis incident photon, with the upper detection plane in blue, and lower plane in green.\\

Moreover, in coded aperture telescopes, the source radiation is spatially modulated on the upper detector by a mask of opaque and transparent elements. The projection of the mask shadow recorded on the upper detection plane produces a shadowgram. This allows simultaneous measurement of source plus background flux (shadowgram area corresponding to the mask holes), and background flux (shadowgram area corresponding to the opaque elements) \citep{Goldwurm_2022}. Knowing the mask pattern, the background can be removed, and the source flux computed, through the deconvolution of the upper detector shadowgram. In this way, a coded mask Compton telescope has the same imaging performances as those of the upper detector.\\

Note that for the regular ISGRI analysis, an iterative process is used to remove the contribution of bright known sources before the deconvolution. Due to the inherent low efficiency of the Compton mode, this bright source cleaning process is not needed with the Compton mode data.

\subsection{Application to IBIS}

\begin{figure}[h!]
    \centering
    \includegraphics[width=.9\linewidth]{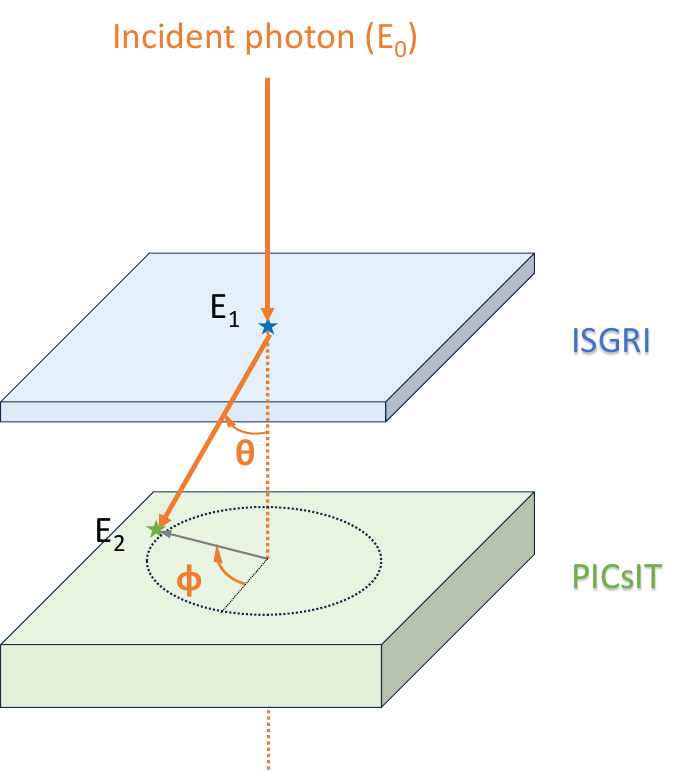}
    \caption{Sketch of a Compton interaction in a Compton telescope. The shown quantities are the initial photon energy ($E_0$), recoil electron energy ($E_1$), scattered photon energy ($E_2$), polar angle ($\theta$), and azimuthal angle ($\phi$).}
    \label{fig:sketch_ibis_compton}
\end{figure}

The IBIS telescope onboard INTEGRAL \citep{Ubertini_2003} can be used as a coded mask Compton telescope thanks to its two detectors (see Fig. \ref{fig:sketch_ibis_compton}):
\begin{itemize}
\item the upper detector: the IBIS Soft Gamma Ray Imager\\ \citep[ISGRI;][]{Lebrun_2003}, made of CdTe semi-conductor pixels  sensitive in the 20 -- 400\,keV energy range ;
\item the lower detector: the Pixellated Imaging Caesium Iodide Telescope \citep[PICsIT;][]{Labanti_2003}, made of CsI(Tl) scintillators sensitive in the 170\,keV -- 10\,MeV energy range.\\
\end{itemize}
The IBIS coded mask element size (11.2 mm), the ISGRI detector pixel size (4.6 mm), and the detector-to-mask distance (3.2 m) result in an angular resolution of \ang{;12;}. The ISGRI-PICsIT interactions are recorded onboard within a 3.6 $\mu$s window, and stored in separate files for on-ground analysis. As stated previously, the IBIS Compton mode, being a coded mask Compton telescope, has the same imaging performances as IBIS/ISGRI.\\

The IBIS Compton mode analysis method was first described by \cite{Forot_2007}, and successfully applied for the first time by \cite{Forot_2008}. This led to many results on bright astrophysical sources, including the Crab Nebula \citep{Forot_2008, Moran_2016}, GRBs \citep{Goetz_2009, Goetz_2013, Goetz_2014}, and Microquasars \citep{Laurent_2011, Rodriguez_2015, Laurent_2016, Cangemi_2023a, Bouchet_2024}. For a short overview of some of those results, see \cite{Goetz_2019}.\\

\subsection{Spectroscopy}

The incident photon energy is estimated by adding both deposited energies, $E_1$ on ISGRI and $E_2$ on PICsIT:
\begin{equation}
    E_0=E_1+E_2
\end{equation}
Then, ISGRI images are computed in a given energy range from IBIS Compton events and deconvolved using standard INTEGRAL/IBIS deconvolution processes (see \cite{Goldwurm_2003} for a description). This allows to find the source count-rate in many energy bands and to build an energy deposit spectrum for each source in the field of view (FOV).\\

\begin{figure}[h!]
    \centering
    \includegraphics[width=\linewidth]{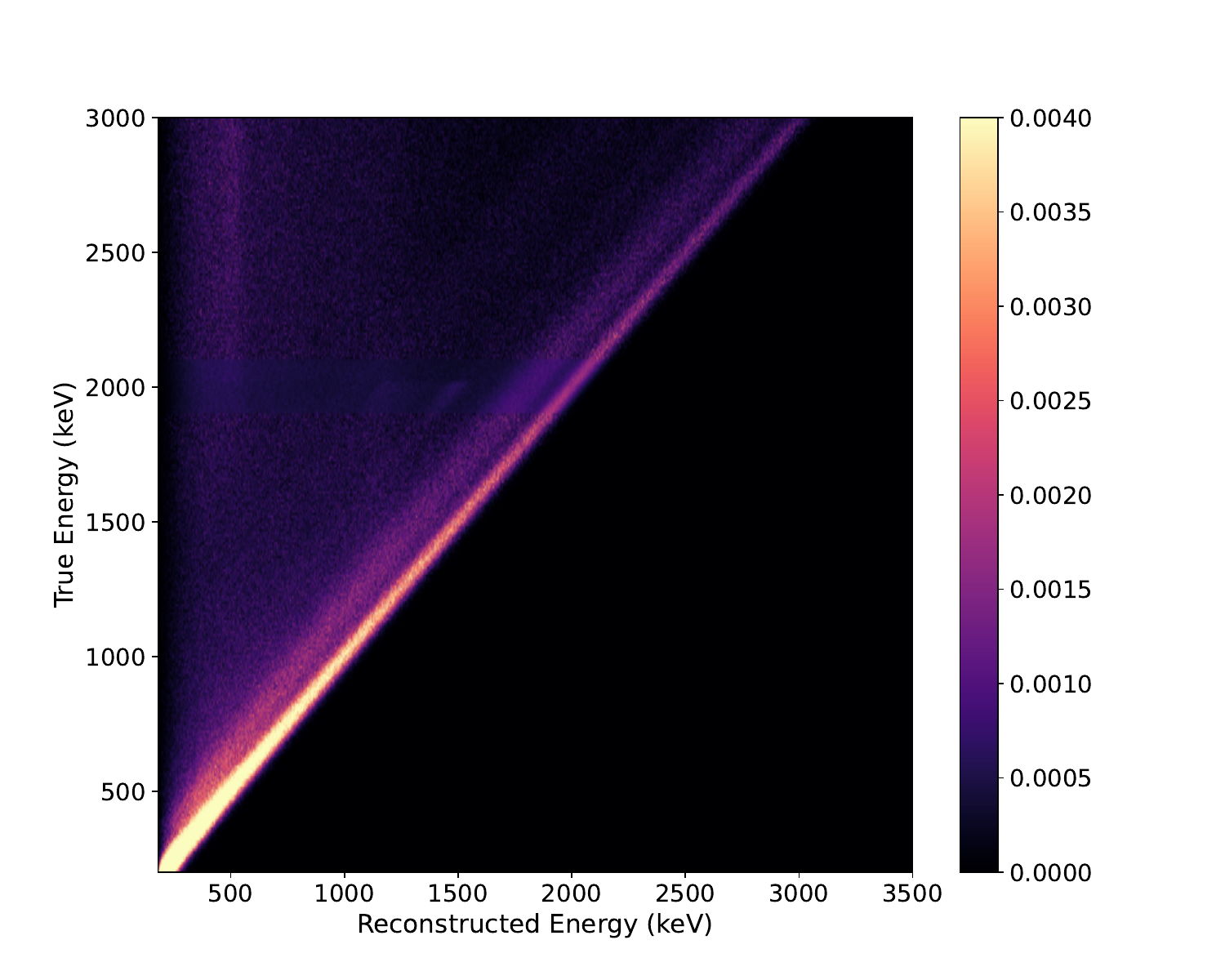}    
    \caption{IBIS/Compton response matrix (RMF), with a saturation value for better contrast.}
    \label{fig:compton_rmf}
\end{figure}

\begin{figure}[h!]
    \centering
    \includegraphics[width=\linewidth]{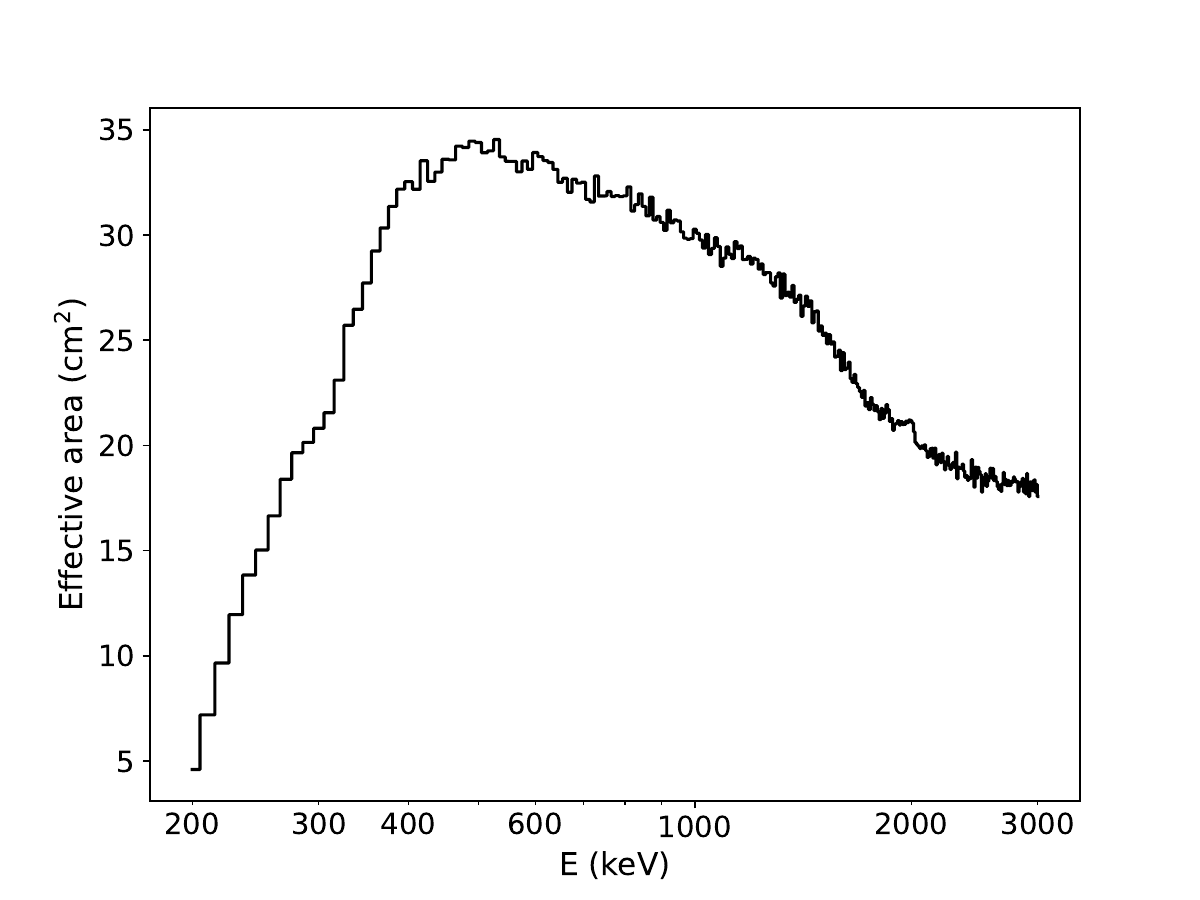}    
    \caption{IBIS/Compton ancillary response (ARF).}
    \label{fig:compton_arf}
\end{figure}

A Geant4 simulation has been developed to simulate the response of the IBIS Compton mode in order to find the real source spectrum. To do so, on-axis monochromatic sources were simulated in 271 channels between 200 -- 3000 keV and the incident Compton photons were recorded between 180 -- 3500 keV across 3321 channels. From this, we built a 3321 x 271 response matrix file (RMF) and a 271 channels ancillary response file (ARF), shown in Fig. \ref{fig:compton_rmf} and \ref{fig:compton_arf} respectively. The spectrum of a model $F_{mod}$ (in photon.s$^{-1}$.cm$^{-2}$) can then be converted into a detector count-rate spectra $C_{mod}$ with:
\begin{equation}
    C_{mod} = (F_{mod}\times\,A)\cdot\,R\cdot\,B =\mathcal{R}(F_{mod})
\end{equation} where $A$ is the ARF, $R$ the RMF, $B$ a boolean re-binning matrix to match the energy channels between the response and spectrum, "$\cdot$" the usual matrix multiplication, and "$\times$" a term by term product. This operation is summarized by the operator $\mathcal{R}$. This model count-rate is then fitted to the energy deposit spectrum determined previously in order to find the "true" source spectrum. 

\subsection{Polarimetry}

While the scattering angle depends on the energy deposits, the azimuthal angle $\phi$ depends, on the other hand, on the polarization (electric field) direction of the source photons. For a single photon with its electric field in the $\Psi_0$ direction, the Compton cross-section depends on the azimuthal angle according to the Klein-Nishina formula:
\begin{equation}
\frac{d\sigma_C}{d\Omega} = \frac{r_e^2}{2} \left(\frac{E_2}{E_0}\right)^2 \left[\frac{E_2}{E_0} + \frac{E_0}{E_2} - 2 \sin^2\theta \cos^2(\phi-\Psi_0) \right]
\end{equation}
where $r_e$ is the classical electron radius, $E_0$ the incident photon energy, $E_2$ the scattered photon energy, and $\theta$ the scattering angle.\\

For a polarized beam of photons, the expected azimuthal angle distribution deduced from the Klein-Nishina formula, called polarigram, is:
\begin{equation} \label{eq:polarigram}
N(\phi)\, d\phi=C\ (1+a_0\cos{(2(\phi-\phi_0))})\,d\phi
\end{equation}
where $N(\phi)$ is the azimuthal angle dependent count-rate in unit of "counts.s$^{-1}$.rad$^{-1}$" -- an "angle-rate" that is independent on the number of angle bins used. $C$ is the mean angle-rate, $a_0$ the amplitude of modulation and $\phi_0$ the preferred azimuthal angle.\\

This distribution is $\pi$-periodic, meaning that we can convert the azimuthal angle of each photon with $\phi_i \longrightarrow \phi_i \mod{\pi}$ to focus only on the $[0,\pi]$ interval. This has the advantage of removing unwanted $2\pi$-periodic modulations in the signal that could appear due to the geometry of the detector, without changing the polarization signature. Moreover, the number of photons per angle is effectively doubled, resulting in higher S/N in each angle bin.\\

In order to measure $N(\phi)$, the Compton events are sorted according to their azimuthal angle, and gathered into $n_p$ polarization bins dividing the interval from 0 to $\pi$ equally. Similarly to what is done for spectra, those distinct event lists are used to build an ISGRI image, which is deconvolved to deduce the source flux for each azimuthal angle bin in a given energy band. The flux are divided by the angle bin size, allowing to build the polarigram, as shown in black in Fig. \ref{fig:example_polarigram}. Note that the data points above \ang{180;;} are simply a copy of the ones below \ang{180;;}, and are only shown for clarity.\\

In fact, this method is even more powerful as it enables independent polarigrams to be derived within a given energy range for each detected source in the FOV. Moreover, choosing the right number of angle bin can be crucial. In general, a value of $n_p=6$ is the minimum required, and is used for lower S/N observation to have a better detection, i.e. a smaller Minimum Detected Polarization (MDP). For high S/N observation, $n_p$ can be increased for a more accurate polarization angle estimation.\\

\begin{figure}[h!]
    \centering
    \includegraphics[width=\linewidth]{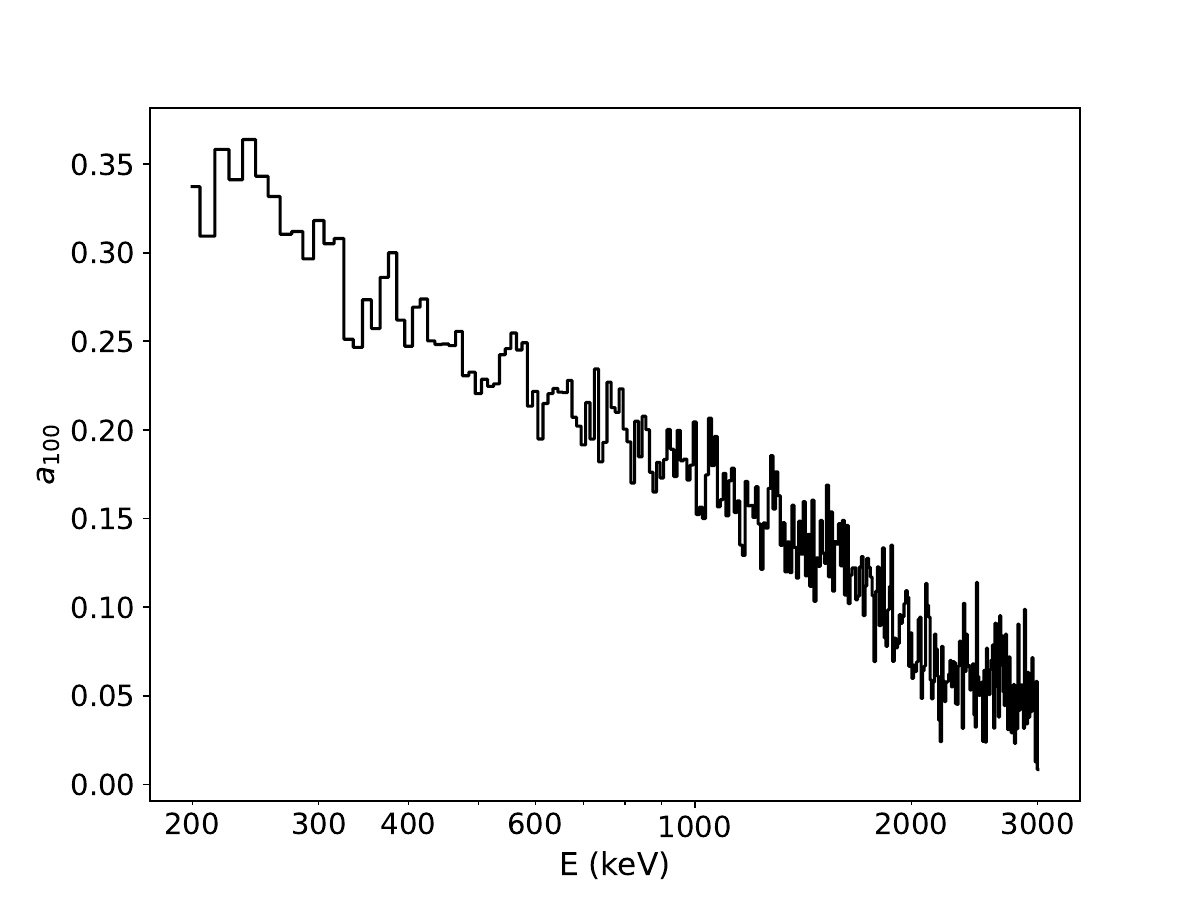}    
    \caption{$a_{100}$ value between 200 and 3000\,keV, determined through Geant4 simulations.}
    \label{fig:a100_200_3000}
\end{figure}

\begin{figure}[h!]
    \centering
    \includegraphics[width=\linewidth]{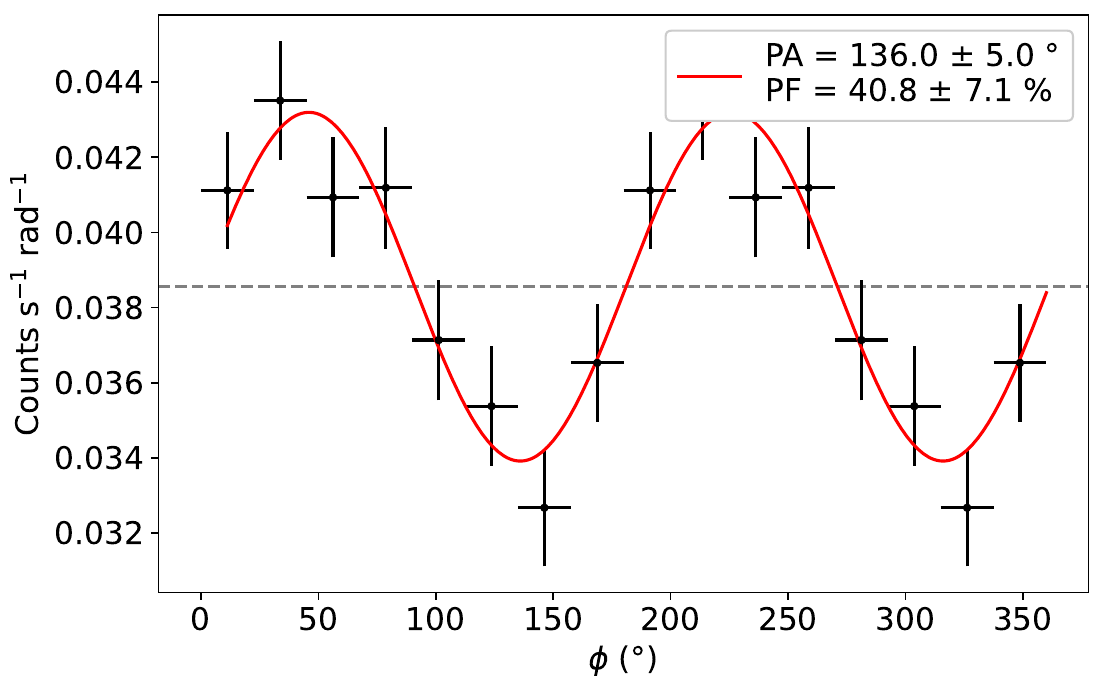}    
    \caption{Example of a polarigram fitted by Eq. \ref{eq:polarigram}. The data used (in black) are from Crab observations in the 200 -- 400\,keV, summed over the 2003--2016 period (see Section \ref{sec:evo_time}).}
    \label{fig:example_polarigram}
\end{figure}

Once the measured polarigram is fitted with Eq. \ref{eq:polarigram}, the Polarization Angle (PA, or $\Psi_0$) is simply deduced with:
\begin{equation}
    \Psi_0=\phi_0+\frac{\pi}{2}\ mod\ \pi
\end{equation}
and the Polarization Fraction\footnote{Sometimes referred to as the Polarization Degree (PD), but we believe this can lead to confusion as the angle are usually given in "degrees".} (PF, or $ \Pi_0$) with:
\begin{equation}
    \Pi_0=a_0/a_{100}
\end{equation}
where $a_{100}$ represents the $a_0$ modulation factor when detecting a 100\% polarized source. Geant4 simulations of a monochromatic, 100\% polarized beam made of 1 million photons allowed us to estimate $a_{100}$ for many energies between 200 and 3000\,keV (see Fig. \ref{fig:a100_200_3000}). The statistical fluctuations are largely due to the low efficiency of Compton scatterings ($\sim 1$\%), which reduces the number of useful photons for the simulation.\\

These values are saved in a Polarization Response File (PRF) and play a similar role as an ARF for spectra. For a specific energy band [E$_1$,E$_2$], we use a flux-weighted average:
\begin{equation}
    a_{100}[E_1,E_2] = \frac{\int_{E_1}^{E_2} a_{100}(E)\,C(E)\, dE}{\int_{E_1}^{E_2}\,C(E)\, dE}
\end{equation}
where $C(E)$ is the Compton rate integrated over all $\phi$. This removes the need to have prior knowledge of the spectral shape.\\

An example of a fitted polarigram using $n_p=8$ is shown in Fig. \ref{fig:example_polarigram}, using data from the Crab Nebula.

\subsection{Statistic of polarization}

Due to the coded mask technique, the individual photon Stokes parameters are not kept during the deconvolution, unlike other instruments such as IXPE; only the polarization parameters -- ($\Pi, \Psi$) -- are deduced from the polarigram. The statistic of those parameters is non-gaussian, as the two variables are correlated. Therefore, estimating the confidence intervals (CI) requires some care. The probability of measuring the parameters ($\Pi, \Psi$) from the observation of a source with a true polarization ($\Pi_s$, $\Psi_s$) is given by the following probability density function (PDF) \citep{Weisskopf_2006, Quinn_2012}:
\begin{equation}\label{eq:polar_pdf}
    P(\Pi,\Psi|\Pi_s,\Psi_s) = \frac{\Pi}{\pi\sigma_0^2} \exp{\left( -\frac{\Pi^2+\Pi_s^2-2\,\Pi\,\Pi_s \cos{(2(\Psi-\Psi_s))}}{2\,\sigma_0^2}\right)}
\end{equation}
where we estimated $\sigma_0^2=\frac{2\sigma_C^2}{n_p\,a_{100}^2C^2}$, following \cite{Weisskopf_2006} and \cite{Laurent_2011}.
To test the null-hypothesis, i.e. the source being unpolarized ($\Pi_s=0$), the PDF is integrated for any angle and above the measured $\Pi=\Pi_0$. The p-value is then:
\begin{equation}\label{eq:p_unpol}
    p =e^{-\ \Pi_0^2/2\sigma_0^2}
\end{equation}
If the p-value is above a set threshold, here $1\%$, the polarization is considered as not detected. An upper-limit is found by inverting Equation \ref{eq:p_unpol} for a chosen $p=p_{lim}$:
\begin{equation}\label{eq:upper_value}
    \Pi_{up}= \sqrt{-2\sigma_0^2\ln{(p_{lim})}} 
\end{equation}
where $p_{lim}=1\%$ hereafter.\\

The uncertainty on the polarization parameters are estimated in the Bayesian framework, from a uniform prior, by marginalizing the posterior probability distribution (see details in Appendix \ref{appendix:marginal}). The best values are defined as the mode, and the uncertainties are found from the Highest Density Interval containing 68\%. The main consequences are to decrease the first estimation of $\Pi_0$ by a few percent, and to give asymmetrical error-bars on $\Pi_0$. Meanwhile, the estimation of $\Psi_0$ remains the same, albeit with larger symmetric error bars. For most measurement in this paper, the detections are significant enough ($\sigma_0^2/\Pi_0^2 \ll 1$) that the upper and lower errors on $\Pi_0$ are approximately equal.\\

\begin{figure}[h!]
    \centering
    \includegraphics[width=\linewidth]{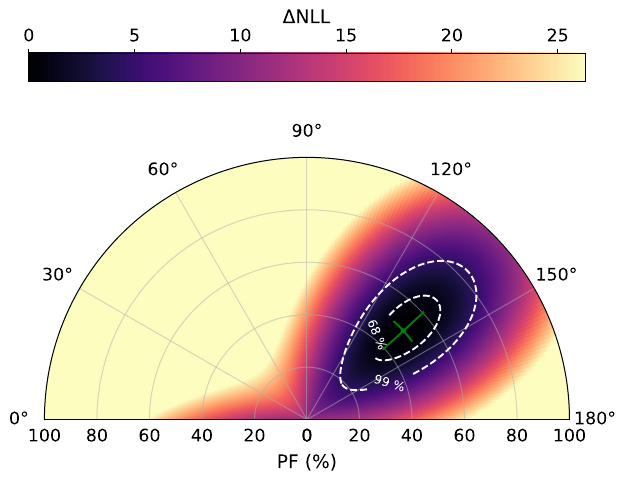}
    \caption{Radar plot of Crab Nebula polarization parameters in the 200 -- 400\,keV band. The best parameters are shown with a green cross. The colormap shows $\Delta$NLL, with a high-cut value used for better contrast, and confidence contours shown with white dotted lines.}
    \label{fig:nll_cn_all_200_400_polar}
\end{figure}

For further statistical assessment, the Delta-Negative Log-likelihood (defined in Appendix \ref{appendix:contours}) can also be computed to find 2D contours at different probability levels. This is shown for the same 2003--2016 period as before in Fig. \ref{fig:nll_cn_all_200_400_polar}.\\ 

\section{The Compton mode analysis}\label{sec:software}

A workflow chart with the overview of the entire Compton IBIS analysis is given in Appendix \ref{app:workflow}, with all the main data-software-simulations interactions. The software is given a list of observation IDs, called SCience Windows (scw), and runs scw by scw from top to bottom for the "Data analysis" part, while the "Simulations" part is ran only once.\\

\subsection{Spurious flux}

Unfortunately, some of the double events are not true Compton scattering and are instead "spurious events", which correspond to simultaneous detection of two independent source and/or background photons \footnote{Two simultaneous background photons is a case already removed through mask deconvolution (see above).}. As we cannot distinguish a true Compton photon from a spurious one, we have to make a global analysis. The main driver of the software is to compute two flux: one from all double events, and one from the spurious events alone using exactly the same analysis scheme. After subtraction, only the true Compton flux from sources in the FOV remain.\\

The double events are already gathered in a file, while the list of "spurious events" is built a posteriori. More precisely, we randomly associate events from the ISGRI and PICsIT single event files. For ISGRI, this event file is directly created onboard, while for PICsIT, such event file is not computed onboard and downloaded to Earth, due to telemetry limitation. Instead, the PICsIT histogram -- a 256-channels spectrum accumulated for each scw and each pixel -- is used to generate a PICsIT event file.\\

As we noted previously, after building this spurious event file, the same analysis is performed for the Compton mode and spurious events, in the desired energy band, eventually for each azimuthal angle bin, to get all Compton and spurious source flux in the FOV. This computation is similar to the standard INTEGRAL Off-line Scientific Analysis (OSA) software  v11.2 \footnote{\url{https://www.isdc.unige.ch/integral/download/osa/doc/11.2/osa_um_ibis/man_html.html}}, with the coded mask deconvolution being performed on the ISGRI detector image. An additional event selection using the Compton scattering angle $\theta$ can be used, but it was not found to change the results significantly and was therefore discarded.\\

The flux obtained from the spurious events has to be multiplied by a "spurious factor" which has remained between 1-3\% throughout the mission. This factor gives the probability of two independent events being flagged as Compton events, and depends on the ISGRI and PICsIT count-rates, as well as the time coincidence window \citep{Forot_2007}. Once this spurious flux $S$ is correctly estimated, we subtract it from the raw Compton flux $C_{r}$ to find the true Compton flux $C_t$:
\begin{equation}\label{eq:compton_flux}
C_t = C_{r} - S
\end{equation}

\subsection{Updated energy calibration}\label{subsec:picsit_energy}

The IBIS Compton mode is quite sensitive to the energy calibration, especially below 400\,keV, where the ARF and PRF vary substantially. Although the OSA 11 analysis software takes the proper calibration of ISGRI into account, it is not the case for PICsIT. To perform a trustworthy Compton analysis we first need to update its calibration. This was specific to our software but could still be applied for the PICsIT single events analysis.\\ 

As the PICsIT CsI(l) scintillator energy conversion is mostly linear \citep{Malaguti_2003}, we used the formula shown below to convert the raw channel number ($c_i$) into energies ($E_i$) for each photon detected by PICsIT:
\begin{equation}\label{eq:picsit_energy}
    E_i\,(keV) = m*gain*(c_i+U_i{[0,1]}) + offset
\end{equation}
where the gain and offset are linear coefficients representing the energy conversion from channel number to photon energy, whose physical origins we do not discuss here. $m$ is a factor that accounts for the different binning between the PICsIT single events (on 256 channels), and PICsIT Compton events (on 64 channels). By definition, $m=1$ for single, and $m=4$ for Compton. $U_i{[0,1]}$ is a random number drawn from a uniform probability between 0 and 1. It avoids underestimating the energy, which on average should sit in the middle of the bin ($c_i+0.5$). It also smooths the spectrum, allowing for a better combination of the spurious and raw Compton, which have different channel sizes. \\

For the first 3 years of the mission, the gain and offset values were deduced from the two radioactive lines of the ${}^{22}\mathrm{Na}$ IBIS calibration source, at 511\,keV and 1275\,keV \cite{Malaguti_2003}. Due to the 2.5 years half-life of the source, we have put in place other methods to estimate those values once those lines were no longer detected by PICsIT. Those methods were found to be stable during the first 3 years of the mission, with reliable results, and we therefore assumed that they should hold for the rest of the mission and ignored possible long-term variations.\\

\subsubsection{PICsIT gain}

\begin{figure}[h!]
    \centering
    \includegraphics[width=\linewidth]{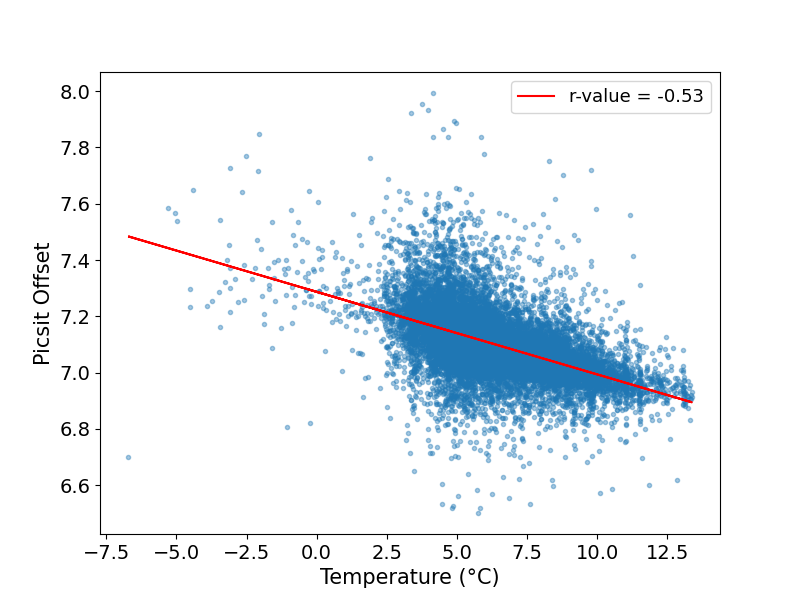}
    \caption{PICsIT detector gain vs temperature correlation. Data is shown as blue dots, and model in red.}
    \label{fig:gain_temp}
\end{figure}

In \cite{Bouchet_2024}, we used a fixed gain of 7.1. This can be improved by using the gain/temperature correlation, as it is well-known that CsI(Tl) energy response is temperature dependent. In particular, it was shown just after launch that the gain is anti-correlated with the detector temperature \cite{Malaguti_2003}. We have thus used this effect to deduce the gain:
\begin{equation}
    gain= a_{T}.T[^{\circ}C] + g_0
\end{equation}
where $a_T$ is the slope coefficient and $g_0$ the gain at $T=0$°\,C. Housekeeping data allow us to monitor the temperature of each module. By averaging the temperature over all module during the course of a scw and comparing with the gain measured with the calibration source, we can find these correlations parameters during the first years of mission. Fig.\ref{fig:gain_temp} shows the correlation with a R-value of $-0.56$, $a_{T}=(2.93 \pm 0.03).10^{-2}$ keV/channel/$^{\circ}$C and $g_0=7.287 \pm 0.002$ keV/channel. The parameters are tightly constrained, as can be seen by the small uncertainties, although there is a high dispersion on $T$, leading to a weak R-value. This is certainly due to the averaging of the temperature over the entire scw and detector.\\

The parameters found imply a gain variation of $\approx0.2\,\%/^{\circ}$C, close to the $0.3\,\%/^{\circ}$C value found in \cite{Malaguti_2003}. The physical origin of this correlation is most likely related to the light yield dependency of the CsI(Tl) crystal with temperature, which is indeed decreasing between -25 and 50°C \citep[see][in particular Fig. 6 and 7]{csi_yield_temperature}.\\

The uncertainty on the gain ($\sigma_g$) is dominated by the dispersion on $T$. We make a rough estimate with:
\begin{equation}
    \sigma_g\approx \sqrt{\frac{1}{N_T} \sum_i \left(g_i-(a_{T}.T_i + g_0)\right)^2} \approx 0.095\,\mathrm{keV/channel}
\end{equation}
with $N_T$ the number of data points, $g_i$ and $T_i$ the data points. For average values of temperature and offset, this results in a dispersion of $\approx5.5$\,keV at 400\,keV. For the polarization, this dispersion is smaller than the energy bin sizes used (> 50\,keV). On the other hand, this could increase the uncertainty on the calibration parameters, which use smaller energy bins (10\,keV). A more precise computation, taking pixel positions and sub-scw timescale into account, is in preparation.\\

\subsubsection{PICsIT offset}

The offset defined in Eq. \ref{eq:picsit_energy} can be computed from the position of a known line; in our case the position of the annihilation line produced by the satellite. Indeed, cosmic rays constantly interact with the various structures of the satellite, inducing activations and leading to the emission of annihilation photons near 511\,keV. This background line can be used in place of the 511\,keV line produced by the $^{22}$Na source to estimate the PICsIT offset. We used PICsIT events recorded as Compton, grouped into histogram during a single scw, to fit this emission line with a gaussian. From the channel position of the line and the already known gain, we recover the offset value.

\subsection{Further event selection}

There is a discrepancy between the single and double events selection onboard. While Compton events are directly transmitted to the On-Board Data Handler unit (OBDH) to be sent on Earth; ISGRI events are first processed, leading to a large portion of them ($\sim$30\%) being discarded before transmission to the ground station. In particular, the single ISGRI events that do not respect certain criteria on the rise time and energy are discarded by the onboard computer to remove false events induced by space protons and to reduce telemetry.\\

To obtain an accurate estimate of the spurious flux, both types of events need to be treated equally in order to be properly combined. We therefore added a post-selection on the Compton events that mimics the onboard computer selection, using the same criteria and parameters \footnote{A detailed explanation can be found in the Tübingen University website at \url{https://uni-tuebingen.de/fakultaeten/mathematisch-naturwissenschaftliche-fakultaet/fachbereiche/physik/institute/astronomie-und-astrophysik/astronomie-hea/forschung/abgeschlossene-projekte/integral-ibis/dokumente/}}.

\section{Calibration with Crab Nebula}\label{sec:calibration}

Checking the accuracy of the Compton mode and correcting any discrepancies require a known calibration source. As stated earlier, there no longer is a radioactive source in IBIS, and we therefore have to rely on a well-known astrophysical source instead: the Crab Nebula.

\subsection{The Crab Nebula, a standard candle for spectroscopy}

The Crab is a bright and relatively stable source at high-energies which is very commonly observed and used as a telescope calibrator. We used Crab INTEGRAL observations with a maximum off-axis angle of \ang{5;;} since the PRF was computed for on-axis events.
Preliminary tests also showed that \ang{5;;} was a good compromise to have minimal off-axis effects on the polarigrams, while maximizing the number of observations in order to have a good S/N \citep{Forot_2008}.\\

\begin{figure}[h!]
    \centering
    \includegraphics[width=\linewidth]{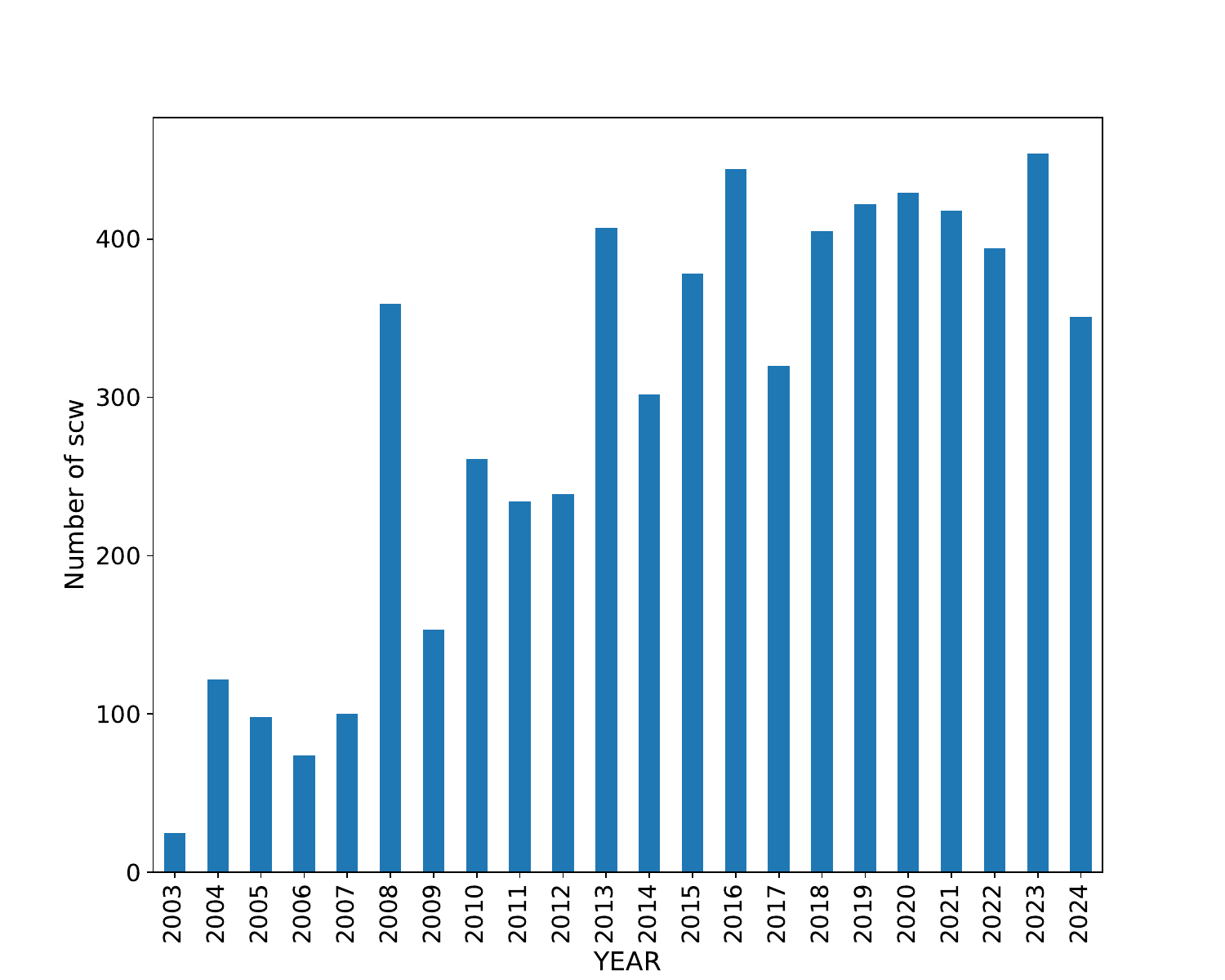}
    \caption{Number of scw with on-axis observation of the Crab Nebula, for each year.}
    \label{fig:crab_scw_distribution}
\end{figure}

INTEGRAL monitored this source regularly during its 22 years of activity (see Fig.\ref{fig:crab_scw_distribution}), amounting to 4663 on-axis scw from 2003-06-01 (MJD\,52791) to 2024-10-11 (MJD\,60594), equating to 10.9\,Ms of exposure. Early observations prior to 2003-06-01 had much bigger coincidence windows, so we ignored them here for consistency.\\

In the soft $\gamma$-rays, the INTEGRAL/SPI instrument \citep{Vedrenne_2003} is the most reliable for spectral comparison, since its calibration and instrumental response are very stable. Indeed, the long term study of the Crab with SPI has shown the light-curve and spectra to be stable over time \citep{Jourdain_2020}, which motivated us to use this source as a canonical spectrum.\\

We relied on the SPI Data Analysis Interface (SPIDAI) \footnote{\url{https://sigma2.irap.omp.eu/integral/spidai}} software to produce a Crab spectrum, which we then fitted using the XSpec software \citep{Arnaud_1996}.
With a power-law model in the 100\,keV -- 1\,MeV range, we find a normalization of $K=18.4\pm4.4$ ph.cm$^{-2}$.s$^{-1}$.keV$^{-1}$ at 1\,keV and photon index of $\Gamma=2.24\pm0.07$ ($\chi^2=24.55$ with 21 dof).\\

Although more sophisticated models can be used -- namely broken power-law and Band model -- these models are only useful when including lower energies. In particular, \citet{Roques_2019} found a photon index $\Gamma_2 = 2.2 - 2.3$ for the higher energy part ($E_{cut} = 100$ keV) of their broken power-law model, in agreement with our fit.\\

\subsection{Estimation of correction parameters}

The simple spurious subtraction estimated from the count-rates on ISGRI and PICsIT single events has shown some limitations over the mission lifetime. In fact, the derived Crab flux suffered from large deviations over time, above the known intrinsic source variation. To correct a posteriori the computation of the true Compton flux, we had to introduce new calibration parameters in Eq. \ref{eq:compton_flux}:
\begin{equation}\label{eq:compton_flux_corr}
C_t = \alpha{(t,E)}\,C_{r} -\beta{(t)}\ S %
,\end{equation}
where $\alpha$ are time and energy dependent parameters that correct the raw Compton rate, and $\beta$ are time dependent parameters that correct the spurious rate. In the High Energy range (HE hereafter, defined as $E>350$\,keV), the $\beta$ correction is sufficient. The correction from $\alpha$ accounts for the discrepancy in the Low Energy range (LE hereafter, defined as $E<350$\,keV), mainly due to the ISGRI low energy threshold evolution with time.\\

All correction parameters are computed by comparing the Crab spectrum obtained through the IBIS Compton mode and the SPI analysis. This standard procedure, used in high-energy telescopes, has been adapted here to correct the systematic error of the spurious flux estimation, since this is a critical step of the analysis.\\

\subsubsection{Spurious correction factor}

\begin{figure}[h!]
    \centering
    \includegraphics[width=\linewidth]{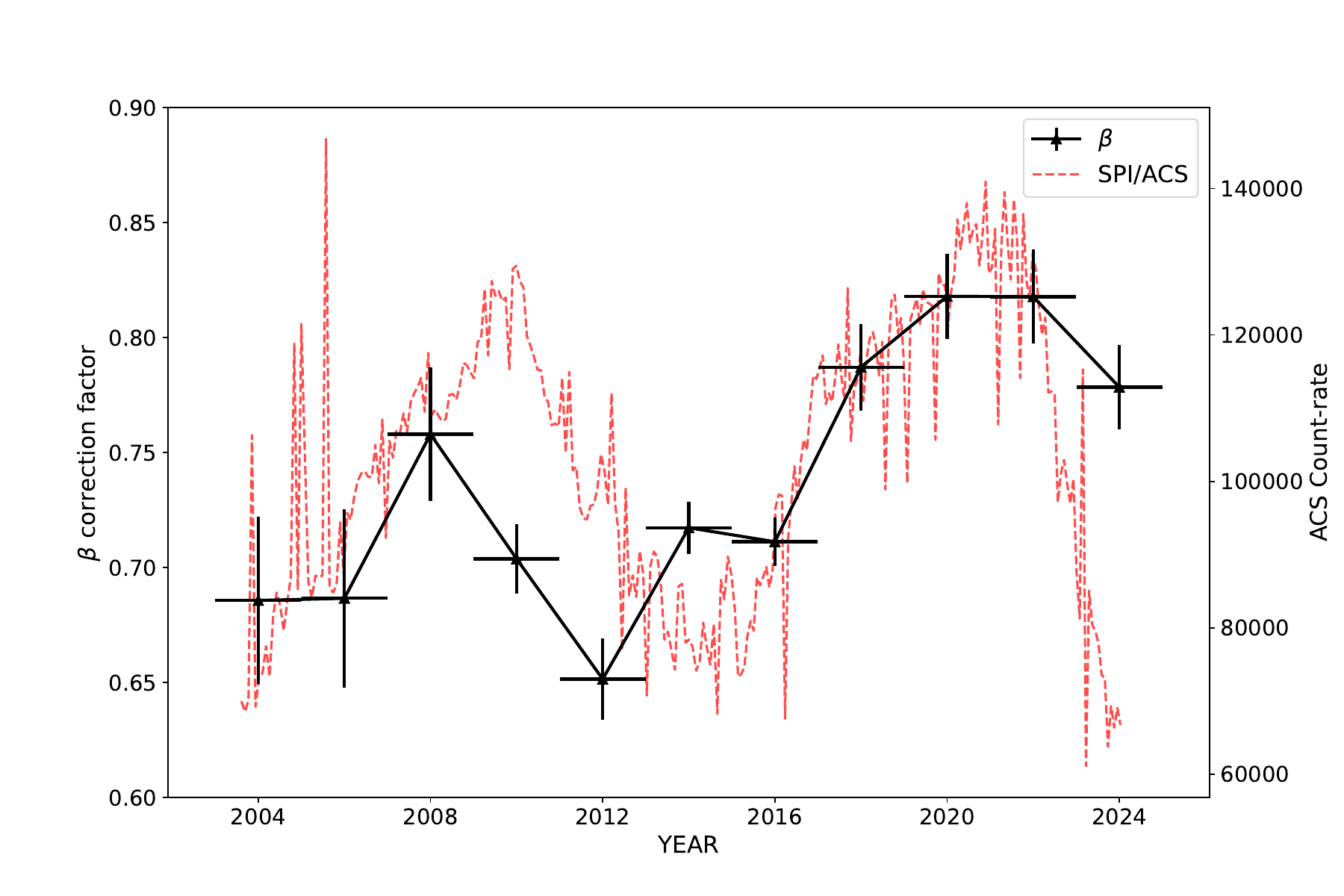}
    \caption{Evolution of the $\beta$ parameter on the left axis (black), compared with the background count-rate on the right axis (SPI/ACS, red).}
    \label{fig:beta_evolution_every2YEAR}
\end{figure}

The first step focuses only on the HE part of the spectrum. To find the right correction, we have to compare the IBIS/Compton and SPI (taken as canonical) spectra of the Crab, and find the value of $\beta$ that minimizes the difference. We used the Levenberg-Marquardt (LM) algorithm, which minimizes the $\chi^2$ function, but instead of finding spectral parameters, the only parameter left free was $\beta$. The SPI flux model is converted into count-rates by the Compton response operator $\mathcal{R}$ and compared with the observed Compton spectrum. The $\chi^2$ is the sum over the energies $E$ of the fit, chosen between 350\,keV and 1.5\,MeV:
\begin{equation}
    \chi^2=\sum_E \frac{(\mathcal{R}(F_{mod})-C_r+\beta S)^2}{\sigma_t^2}
\end{equation}
The method is repeated for different periods, and the LM algorithm finds the best-fit $\beta$ for each. We choose a duration of 2 years for each period, as this allows for small enough statistical errors, while following the variations with time. The resulting evolution is shown in Fig. \ref{fig:beta_evolution_every2YEAR}, where $\beta$ varies between 0.65 and 0.85, roughly following the cosmic ray background of the satellite (in red), that is measured from the Anti Coincidence Shield (ACS) of SPI. The uncertainty on $\beta$, $\sigma_{\beta}$, is found using the standard least-square uncertainties evaluation from the covariance matrix.

\subsubsection{Raw Compton correction factor}

\begin{figure}[h!]
    \centering
    \includegraphics[width=\linewidth]{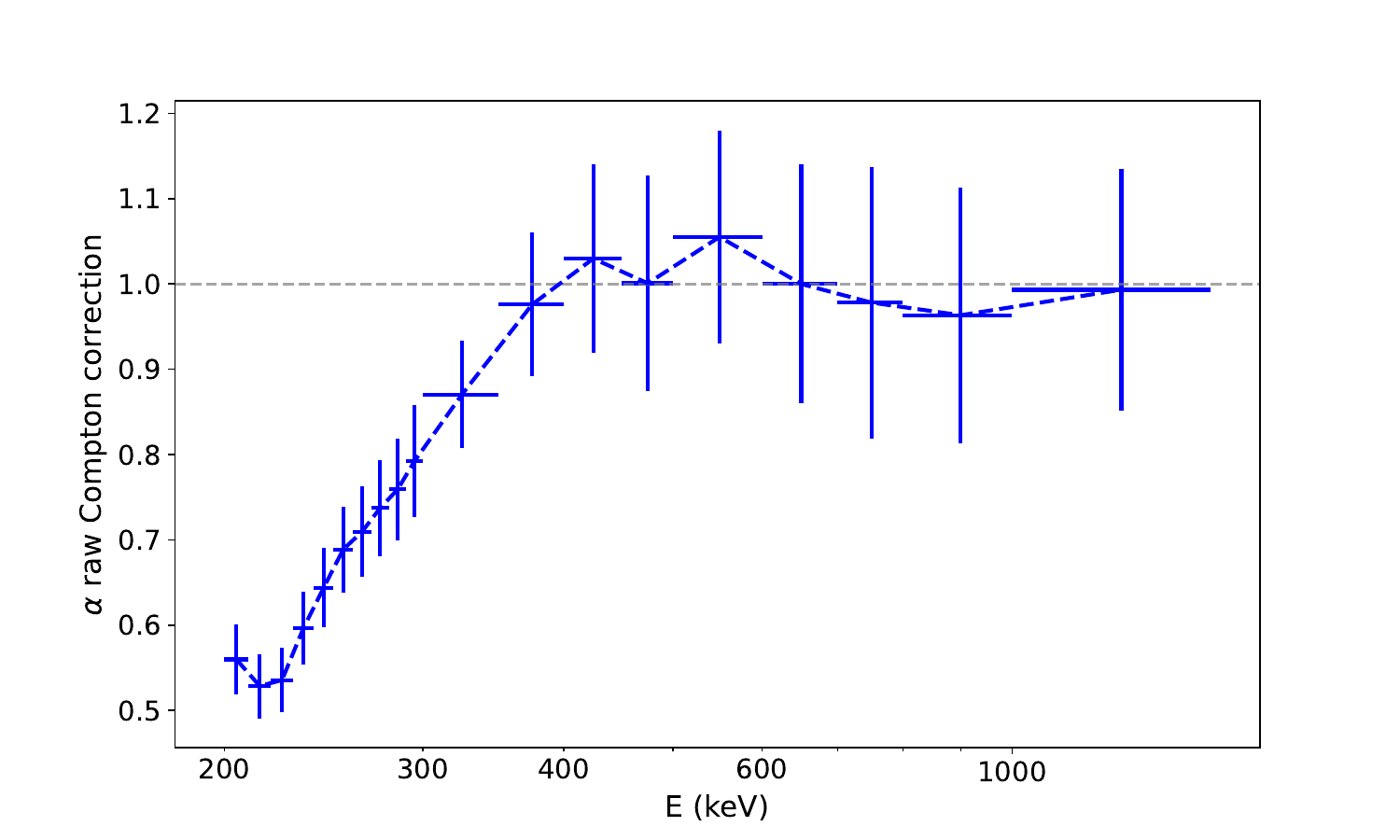}
    \caption{Evolution of the $\alpha$ parameter with energy for the 2013--2015 period. The value $\alpha=1$ is shown as a grey-dotted line.}
    \label{fig:alpha_raw_correction_2013_2015}
\end{figure}

\begin{figure}[h!]
    \centering
    \includegraphics[width=\linewidth]{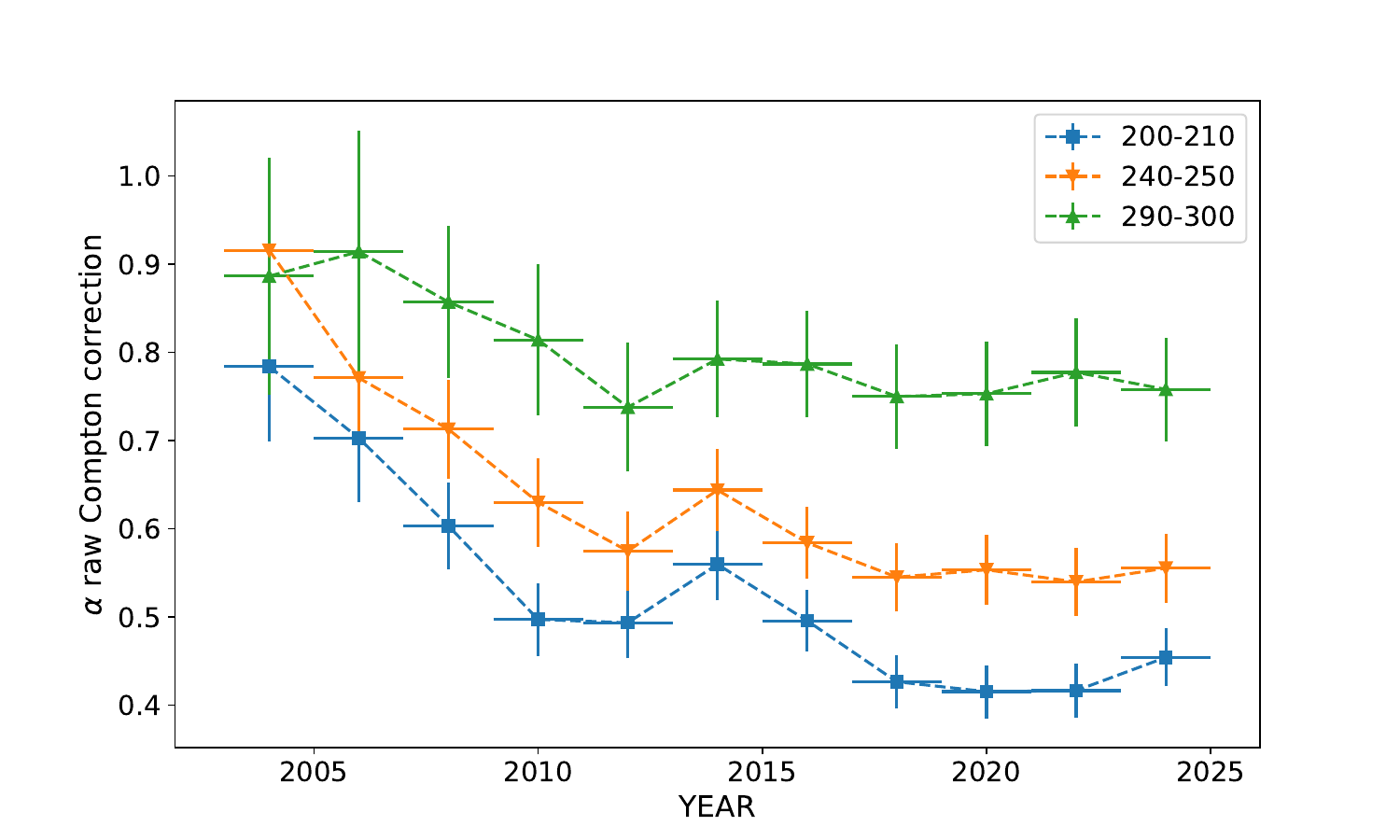}
    \caption{Evolution of the $\alpha$ parameter with time for the first energy bands.}
    \label{fig:alpha_evolution_every2YEAR}
\end{figure}

Once this first correction is achieved, we have to compute the $\alpha$ parameters for the entire energy range. The LE range in particular requires an important correction, which could not be achieved with the $\beta$ parameters alone. We used Eq.\ref{eq:compton_flux_corr} to isolate $\alpha$, and used the SPI model as a reference by setting $C_t=\mathcal{R}(F_{mod})$. For a specific period, we have:
\begin{equation}\label{eq:alpha_corr}
    \alpha(E)= \frac{C_{mod}+\beta\,S}{C_r}
\end{equation}
The uncertainty on $\alpha$ is found by differentiating Eq. \ref{eq:alpha_corr} with respect to the raw Compton rate and corrected spurious rate ($\beta S$), and propagating the errors to $\alpha$. For the 2013--2015 period, the evolution with energy is shown in Fig. \ref{fig:alpha_raw_correction_2013_2015}. As expected, $\beta$ alone is sufficient above 350\,keV, and therefore $\alpha$ stays close to 1 with less than $\pm$0.1 amplitude of variation. In the LE, $\alpha$ is well below 1, and thus compensates for the excess of flux compared to the model. The variation with time of some energy bands are shown in Fig. \ref{fig:alpha_evolution_every2YEAR}. All parameters seem to decrease until 2010, when it becomes relatively stable. The physical origins of these deviations in energy and time is probably linked to the ISGRI low threshold evolution, which is still under investigation.\\

Both $\alpha$ and $\beta$ parameters are computed once and saved in a table, so corrected flux can be computed during the last step of the Compton mode analysis. The errors are propagated as follows:
\begin{equation}\label{eq:compton_errors}
\sigma_t^2 = \alpha^2\,\sigma_{r}^2 + \beta^2\sigma_S^2 + C_r^2 \sigma_{\alpha}^2 + S^2 \sigma_{\beta}^2
,\end{equation}
The error on the final flux is dominated by the error terms from $\sigma_{r}$ and $\sigma_{S}$; while the error on $\alpha$ and $\beta$ have overall negligible impact. We note that the uncertainty on the PICsIT energies (Section \ref{subsec:picsit_energy}) could also lead to underestimated errors.

\subsection{Compton spectrum}

\begin{figure}[h!]
    \centering
    \includegraphics[width=\linewidth]{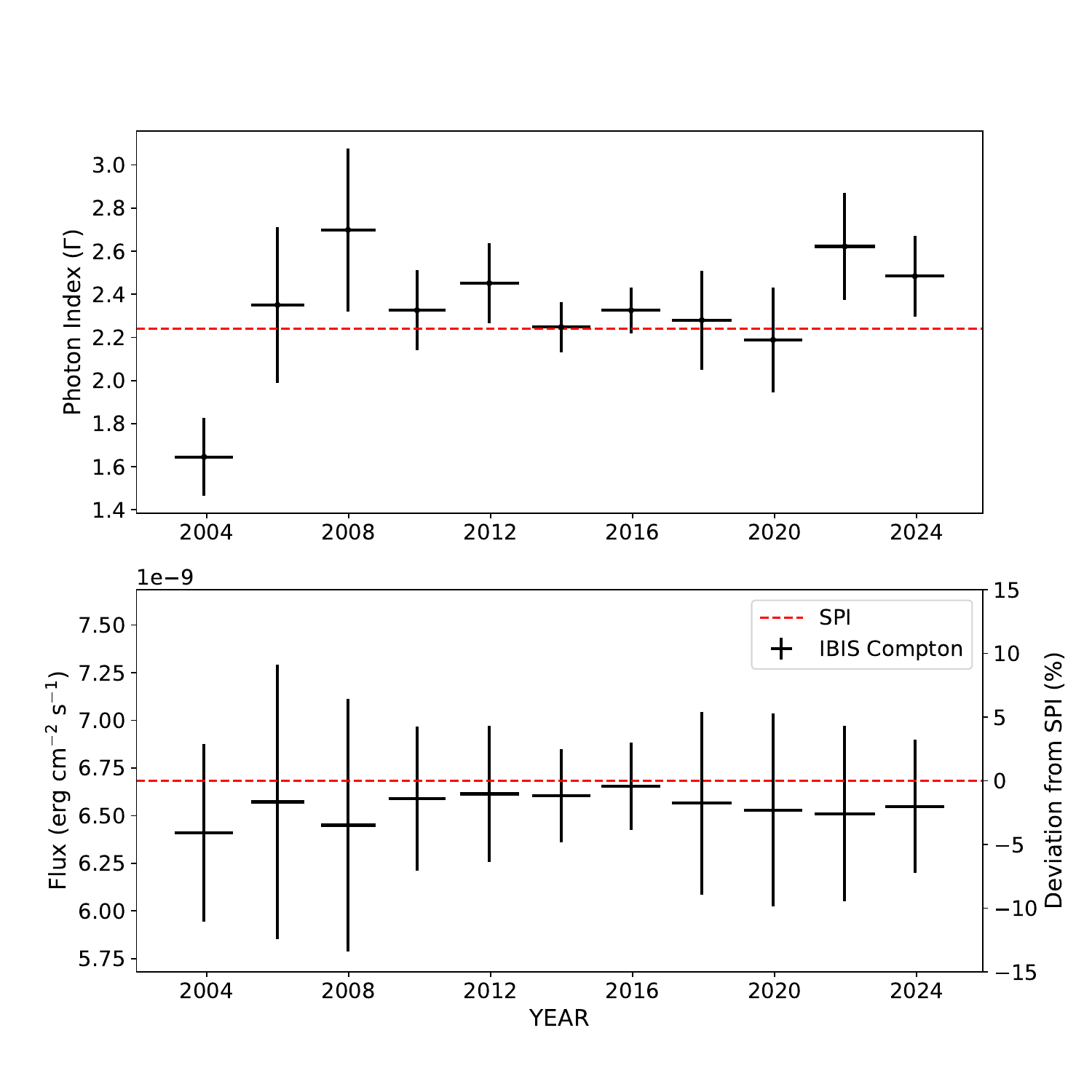}
    \caption{Evolution of the Crab spectral parameters from 300 -- 1500\,keV fits. The flux is defined in the 350 -- 1000\,keV range.}
    \label{fig:crab_gamma_flux_2YEAR}
\end{figure}

To check the $\beta$ correction, Compton mode spectra are built every two years and fitted with a  power-law between 350 and 1500\,keV. In a standard power-law model, the $K$ normalization factor is defined at 1\,keV, which is hard to interpret when fitting spectra two orders of magnitude above.
Instead, we used the flux in the 350 -- 1000\,keV band, $F[E_1,E_2]$, as the normalization. The power-law is re-written:
\begin{equation}
    N(E)=F[E_1,E_2]\,\frac{(\Gamma-2)E^{-\Gamma}}{(E_1^{2-\Gamma}-E_2^{2-\Gamma})}
\end{equation}
where $E_1=350$\,keV and $E_2=1000$\,keV. The evolution of the spectral parameters is shown in Fig. \ref{fig:crab_gamma_flux_2YEAR} compared to SPI values and in Table \ref{tab:crab_spec_param}. Some deviation of $\Gamma$ is seen at the start of the mission, when the telescope was not yet well calibrated, as well as in the last 4 years, a behavior still under study. The flux remain on the other hand compatible throughout.\\

\begin{figure}[h!]
    \centering
    \includegraphics[width=\linewidth]{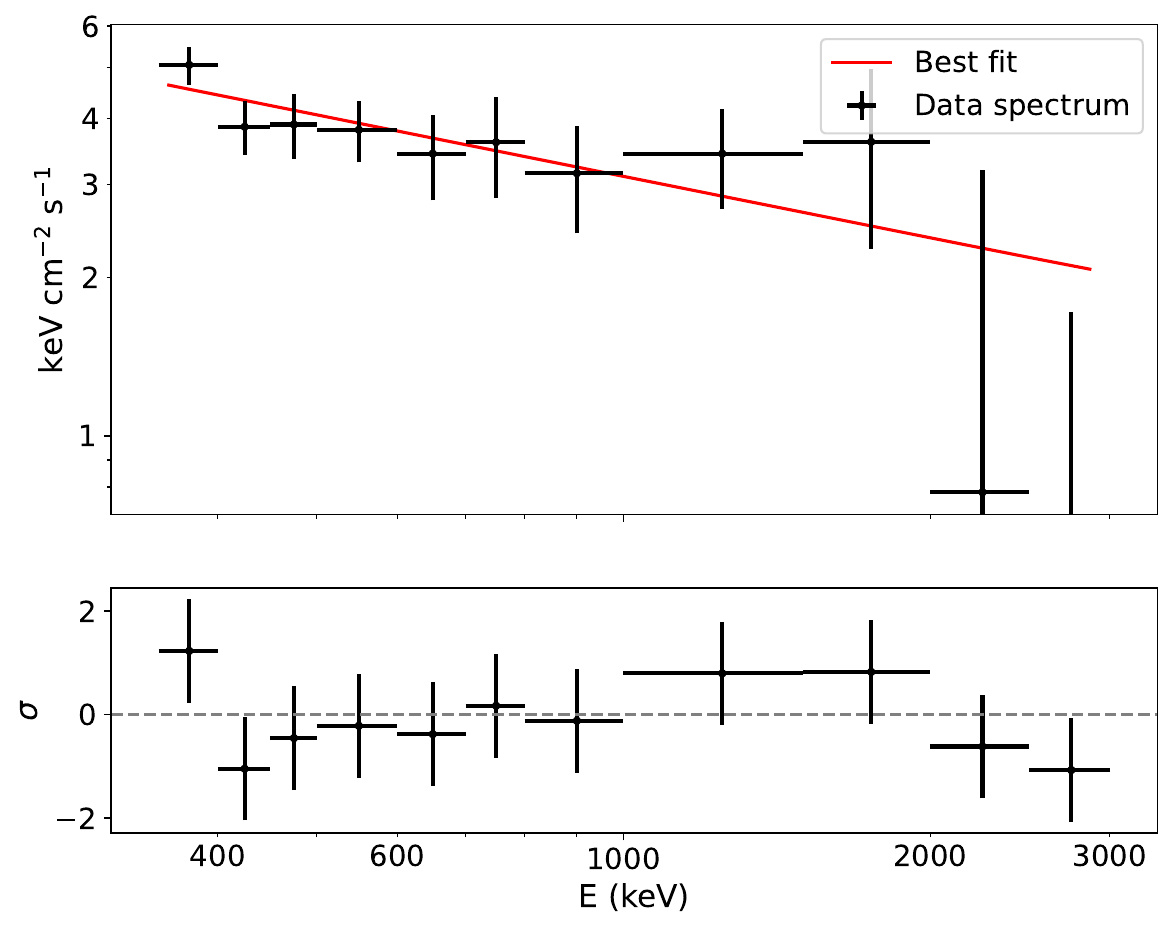}
    \caption{Spectral fit of combined Crab observations in the 2003 - 2024 period (in red), in the 0.35 -- 3\,MeV range. Note that the data points flux shown in black are estimated using the "flux/count-rate" ratio of the model.}
    \label{fig:crab_fit_all_350_3000}
\end{figure}

A spectrum of the combined dataset is also shown up to 3\,MeV on Fig. \ref{fig:crab_fit_all_350_3000}. Using the previous model, we find $\Gamma=2.28 \pm 0.11$ and a 350 -- 1000\,keV flux of $(6.46\pm0.27)\,10^{-9}$erg.s$^{-1}$.cm$^{-2}$, in good agreement with the SPI model. Those results thus confirm that the Compton mode spectral calibration is sound, and can be further used for polarization analysis.

\section{Results: Crab Nebula polarization}\label{sec:polar}

The Crab is the source whose polarization above 1\,keV has been measured by the most instruments (see Table \ref{tab:crab_polar_others}), followed by Cygnus\,X-1. The PAs are grouped near the pulsar rotation axis at $\Psi_{CP}=124.0\pm0.1$° \citep{crab_pulsar_axis}, within roughly $\pm20$°. The PF is measured above 15\% and up to 47\%, confirming the synchrotron origin of the emission between 1\,keV and 1\,MeV.

\begin{table}[h!]
\caption{Summary of 21st century high-energy polarization measurements of the Crab, in chronological order of observation.}
\label{tab:crab_polar_others}
\centering
\begin{threeparttable}
{\renewcommand{\arraystretch}{1.3}
\begin{tabular}{cccc}
Instrument & Energy (keV)  & PA (°) & PF (\%) \\
    \hline
    \midrule
SPI$^1$ & 130 -- 436 & 120 $\pm$ 6 & 24 $\pm$ 4\\
IBIS$^2$ & 200 -- 800  & 100 $\pm$ 11 & 47 $\pm$ 13\\
Hitomi/SGD$^3$ & 60 -- 160  & 110 $\pm$ 13& 22$ \pm$ 10\\
AstroSat/CZTI$^4$ & 100 -- 380  & 143.5 $\pm$ 2.8 & 39 $\pm$ 10\\
PoGo+$^5$ & 18 -- 160 & 131.3 $\pm$ 6.8 &  20.9 $\pm$ 5.0\\
PolarLight$^6$ & 3 -- 4.5 & 145.8 $\pm$ 5.7 & 15.3 $\pm$ 3.1 \\
IXPE$^7$ & 2 -- 8 & 145.5 $\pm$ 0.29 & 19.0 $\pm$ 0.2\\
XL-Calibur$^8$ & 19 -- 64 & 129.8 $\pm$ 3.2° & 25.1 $\pm$ 2.9 \\
\hline
\end{tabular}}
    \begin{tablenotes}
    \small
    \item $^1$\cite{Dean_2008}, revisited by \cite{Jourdain_2019}; $^2$\cite{Forot_2008}; $^3$\cite{hitomi_2018}; $^4$\cite{Vadawale_2017}; $^5$\cite{Chauvin_2018}; $^6$\cite{polar_light_crab}; $^7$\cite{ixpe_crab_polar}; $^8$\cite{xcalibur_crab}
    \end{tablenotes}
    \end{threeparttable}
\end{table}

Despite those large deviations between instruments, we used these values for comparison to our results, keeping in mind that large systematic errors probably exist for some of those experiments.

\subsection{Combined data set}

\begin{figure}[h!]
    \centering
    \includegraphics[width=\linewidth]{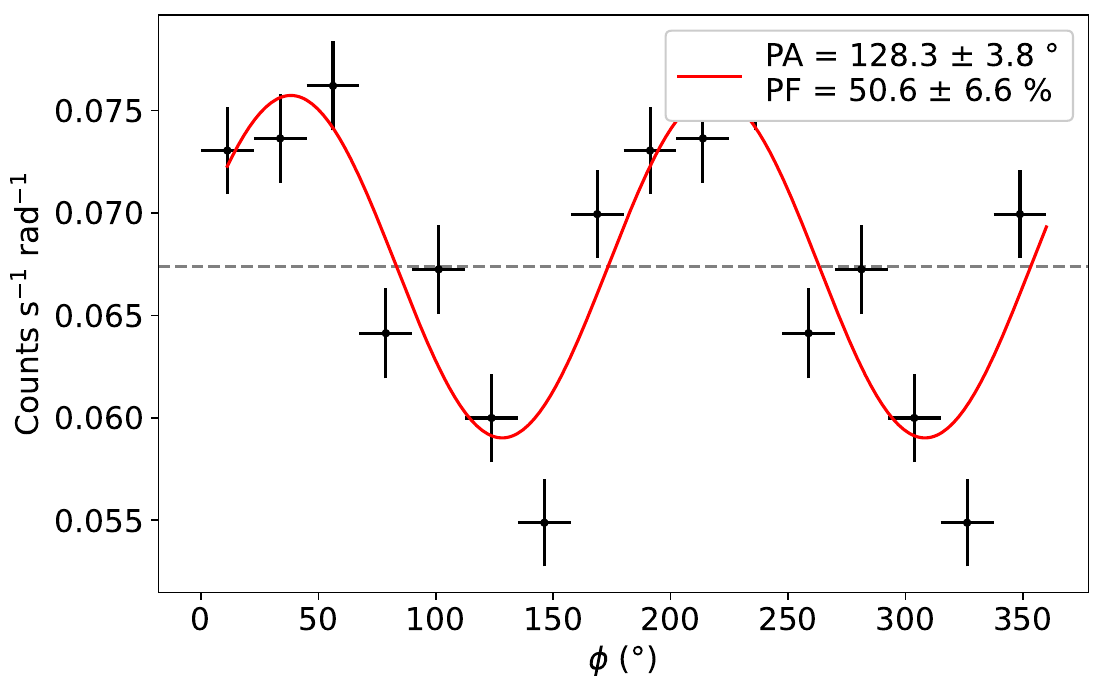}
    \caption{Polarigram of all combined data of the Crab Nebula in the 200 -- 1000\,keV range.}
    \label{fig:polarigram_200_1000_8p}
\end{figure}

We used $n_p=8$ for the azimuthal angle binning, as it allowed to have small uncertainties on polarization parameters, while still maintaining a good p-value. Combining all data from 2003-06-01 to 2025-01-01 between 200 -- 1000\,keV results in the polarigram shown in Fig. \ref{fig:polarigram_200_1000_8p} in black. After fitting and applying the statistical analysis described in Section \ref{sec:polar}, we find a polarization angle of 128.3$\pm$3.8° and a fraction of 50.6$\pm$6.6 \%.\\

\begin{figure}[h!]
    \centering
    \includegraphics[width=\linewidth]{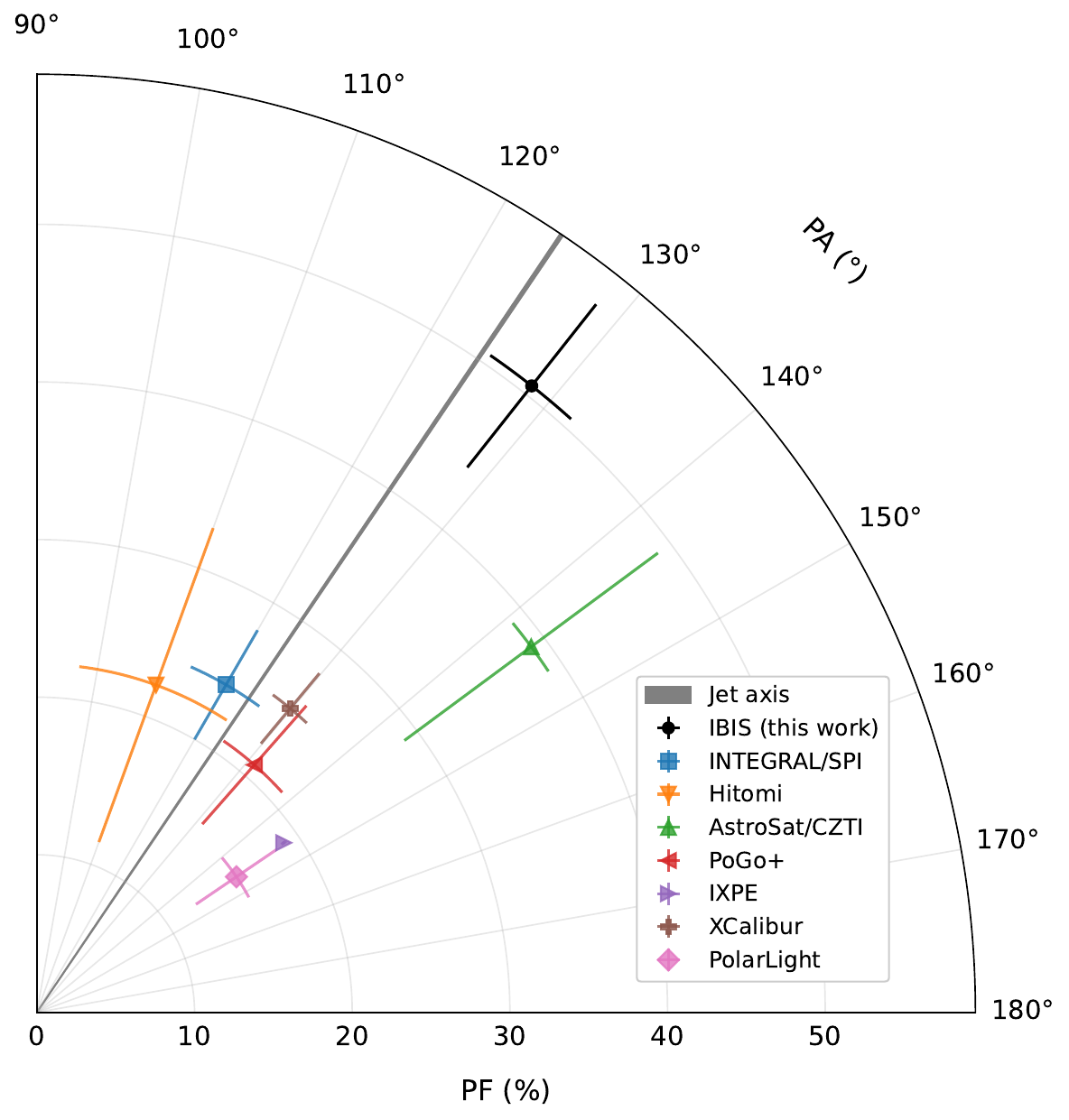}
    \caption{Radar plot of the Crab Nebula polarization parameters measured with the IBIS Compton mode in the 200 -- 1000\,keV band (black). Other measurements from Table \ref{tab:crab_polar_others} are shown as: INTEGRAL/SPI (blue), AstroSat (green), Hitomi (orange), PoGO+ (red), IXPE (purple), XL-Calibur (brown), and PolarLight (pink). The Crab Pulsar axis position angle is indicated by a gray line.}
    \label{fig:crab_quarter_radar_200_1000_kev}
\end{figure}

This result is shown in black in Fig. \ref{fig:crab_quarter_radar_200_1000_kev}, compared with other experiments (in colors). The PA is close to the Crab Pulsar axis (in gray), while the PF is much higher than the other measurements, being only compatible with AstroSat. This is likely due to the energy band considered here, which encompasses higher energies than the other experiments, with only AstroSat and SPI having comparable energies. As the Crab polarization has been shown to change with time \citep{Moran_2016}, we have also studied its evolution.

\subsection{Crab Nebula polarization: time evolution}\label{sec:evo_time}

The Crab dataset was cut into 6 periods of roughly equal exposures, in order to obtain enough S/N for a detection. We considered only the 200 -- 400\,keV band, as including higher energy bands does not improve the detection. The results are shown in Table \ref{tab:crab_timepolar}, along with the p-values.\\

\begin{figure}[h!]
    \centering
    \includegraphics[width=.9\linewidth]{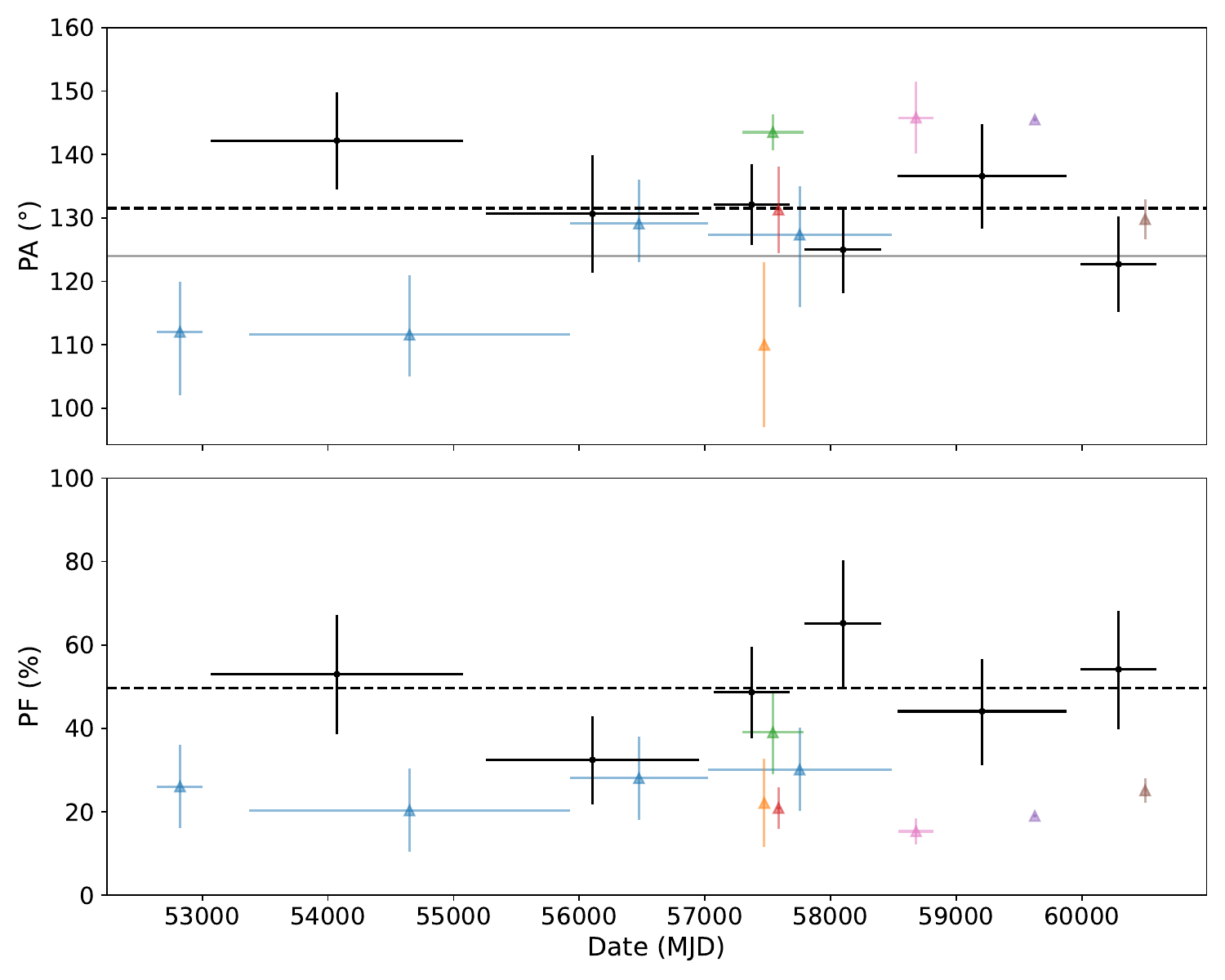}
    \caption{Evolution of Crab polarization measured by IBIS in the 200 -- 400\,keV range with the average values shown as black dotted lines, and Crab Pulsar axis in gray. Other measurements from Table \ref{tab:crab_polar_others} are shown as: INTEGRAL/SPI (blue), AstroSat (green), Hitomi (orange), PoGO+ (red), IXPE (purple), XL-Calibur (brown), and PolarLight (pink).}
    \label{fig:crab_timepolar_200_400_kev}
\end{figure}

A visual comparison with other missions is shown in Fig. \ref{fig:crab_timepolar_200_400_kev}. The PA is compatible with most missions, even following the recent variation in angle after MJD\,60000 seen by XL-Calibur. The main discrepancy seen is with the SPI measurement before MJD\,55000 (in blue), in the early phases of the INTEGRAL mission. This could be due to the calibration of the instruments, which was not settled in this early period. The PF seems larger than most other missions, with only SPI and AstroSat (in green) being compatible. This is probably a difference intrinsic to the source, as the incompatible measurements are all from lower energies (see Table \ref{tab:crab_polar_others}). This is corroborated by the fact that the Crab spectrum has a known break around 100\,keV \citep{Jourdain_2020}, which could be correlated with a change in polarization behavior. Finally, the variation in time can partially explain the discrepancy seen for the combined data set.

\begin{table*}[h!]
    \caption{Crab polarization for different periods measured by the IBIS Compton mode in the 200 -- 400\,keV range.}
    \centering
    \begin{tabular}{ccccccc}
    MJD Start & MJD End & EXPO (ks) & S/N & PA (°) & PF (\%) & p-value (\%) \\
    \hline
    \midrule
    53069.8 & 55072.5 & 1395 & 35 & 142.2 $\pm$ 7.7 & 53 $\pm$ 14 & 0.05 \\
    55258.8 & 56952.1 & 2032 & 48 & 130.6 $\pm$ 9.3 & 32 $\pm$ 11 & 0.4 \\
    57068.6 & 57675.3 & 1619 & 45 & 132.1 $\pm$ 6.5 & 49 $\pm$ 11 & 0.003 \\
    57794.6 & 58404.0 & 1412 & 32 & 125.0 $\pm$ 6.9 & 65 $\pm$ 15 & 0.007 \\
    58531.2 & 59880.5 & 3027 & 39 & 136.6 $\pm$ 8.3 & 44 $\pm$ 13 & 0.1 \\
    59989.4 & 60594.6 & 1412 & 35 & 122.7 $\pm$ 7.5 & 54 $\pm$ 14 & 0.03 \\
    \hline
    \end{tabular}
    \label{tab:crab_timepolar}
\end{table*}

The previous Crab Nebula study done by \cite{Moran_2016} with IBIS data showed a quite different picture. Their polarization angle in the 300 -- 450\,keV range varied from 115$\pm$11° to 80$\pm$12° between the two time periods of 2003--2007 and 2012--2014. Those results were obtained with a different energy calibration, and included higher off-axis angle observations. A future study including the phase-resolved Crab Nebula polarization, as done by \cite{Forot_2008}, could shed light on the nature of those differences, and whether they are of instrumental origin.\\

Overall, the established relative stability of the Crab spectrum over time \citep{Jourdain_2020} seems to conflict with the large variations of the polarization. One explanation would be that only the orientation of the magnetic field at the origin of the synchrotron emission varies, while its strength remains unchanged.

\subsection{Evolution with energy}

\begin{figure*}[h!]
    \centering
    \includegraphics[width=.9\linewidth]{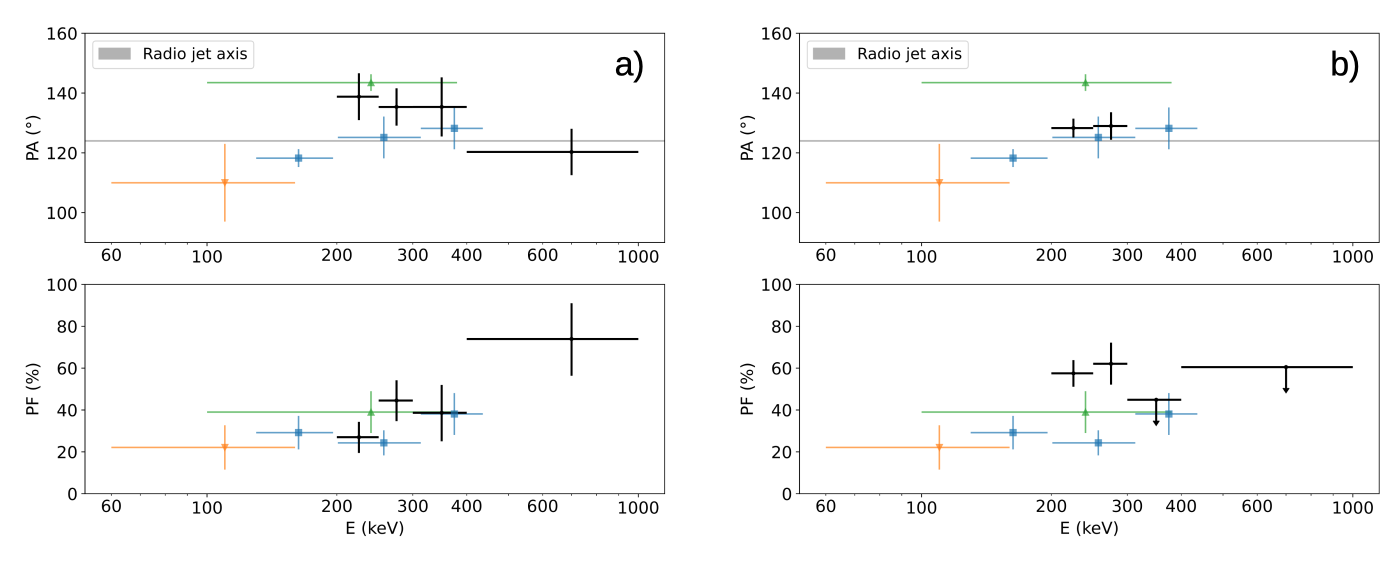}
    \caption{Evolution of the Crab Nebula polarization with energy in black. \textbf{a)} 2003--2016 period. \textbf{b)} 2016--2024 period. The other missions shown are INTEGRAL/SPI (blue), AstroSat (green) and Hitomi (orange).}
    \label{fig:crab_specpolar_both_periods}
\end{figure*}

To explore further those discrepancies, we made a study of the polarization evolution with energy. We separated the dataset in two periods of similar exposures and time span: 2003-06-01 to 2016-06-01 (Period\,1), and 2016-06-01 to 2025-01-01 (Period\,2), with 4.6\,Ms and 6.2\,Ms respectively. Four energy bands were chosen to keep a good S/N (200--250, 250--300, 300--400, and 400--1000\,keV) and we only kept other experiments above 100\,keV for comparison. Results for both periods are summarized in Table \ref{tab:crab_specpolar_periods} and shown in Fig. \ref{fig:crab_specpolar_both_periods}.\\

For Period 1 (left side), polarization is detected in all the energy bands. The PA is nearly constant up to 400\,keV, and shows a hint of a decrease by $\sim$\ang{20;;} above 400\,keV, but the measurements are still marginally compatible within the errors at 68\%. The PF increases with energy, although the errors bars are consistent with each other for the energy bands below 400\,keV.\\

For Period 2, results are shown in Fig. \ref{fig:crab_specpolar_both_periods} (right). Polarization is detected only up to 300\,keV, with upper-limits above 40\% for the PF. The PAs between 200 -- 300\,keV are compatible with the previous period, while the PF is quite high and decreases between the 200 -- 250 and 250 -- 300\,keV band.\\

\begin{table*}[h!]
    \centering
    \caption{Crab polarization for the main two periods measured by the IBIS Compton mode.}
    \begin{tabular}{ccccc|cccc}
    E (keV) & S/N & PA (°) & PF (\%) & p-value (\%) & S/N & PA (°) & PF (\%) & p-value (\%) \\
    \hhline{=====|====}
    & & & & & & & & \\
    & \multicolumn{4}{l|}{\textit{Period 1:} MJD 52866--57494 (Expo. = 4681 ks)} & \multicolumn{4}{l}{\textit{Period 2:} MJD 57619--60595 (Expo. = 6257 ks)} \\
    & & & & & & & & \\
    200 -- 250 & 57 & 138.8 $\pm$ 7.9 & 27.0 $\pm$ 7.4 & 0.06 & 65 & 128.3 $\pm$ 3.2 & 57.5 $\pm$ 6.4 & 1e-16 \\
    250 -- 300 & 45 & 135.3 $\pm$ 6.3 & 44.5 $\pm$ 9.8 & 0.002 & 43 & 129.0 $\pm$ 4.7 & 62 $\pm$ 10 & 3e-7 \\
    300 -- 400 & 39 & 135.3 $\pm$ 9.9 & 38 $\pm$ 13 & 0.7 & 34 & - & < 45 & 5 \\
    400 -- 1000 & 32 & 120.3 $\pm$ 7.8 & 73 $\pm$ 17 & 0.04 & 31 & - & < 60 & 50 \\
    \hline
    \end{tabular}
    \label{tab:crab_specpolar_periods}
\end{table*}

\begin{figure}[h!]
    \centering
    \includegraphics[width=\linewidth]{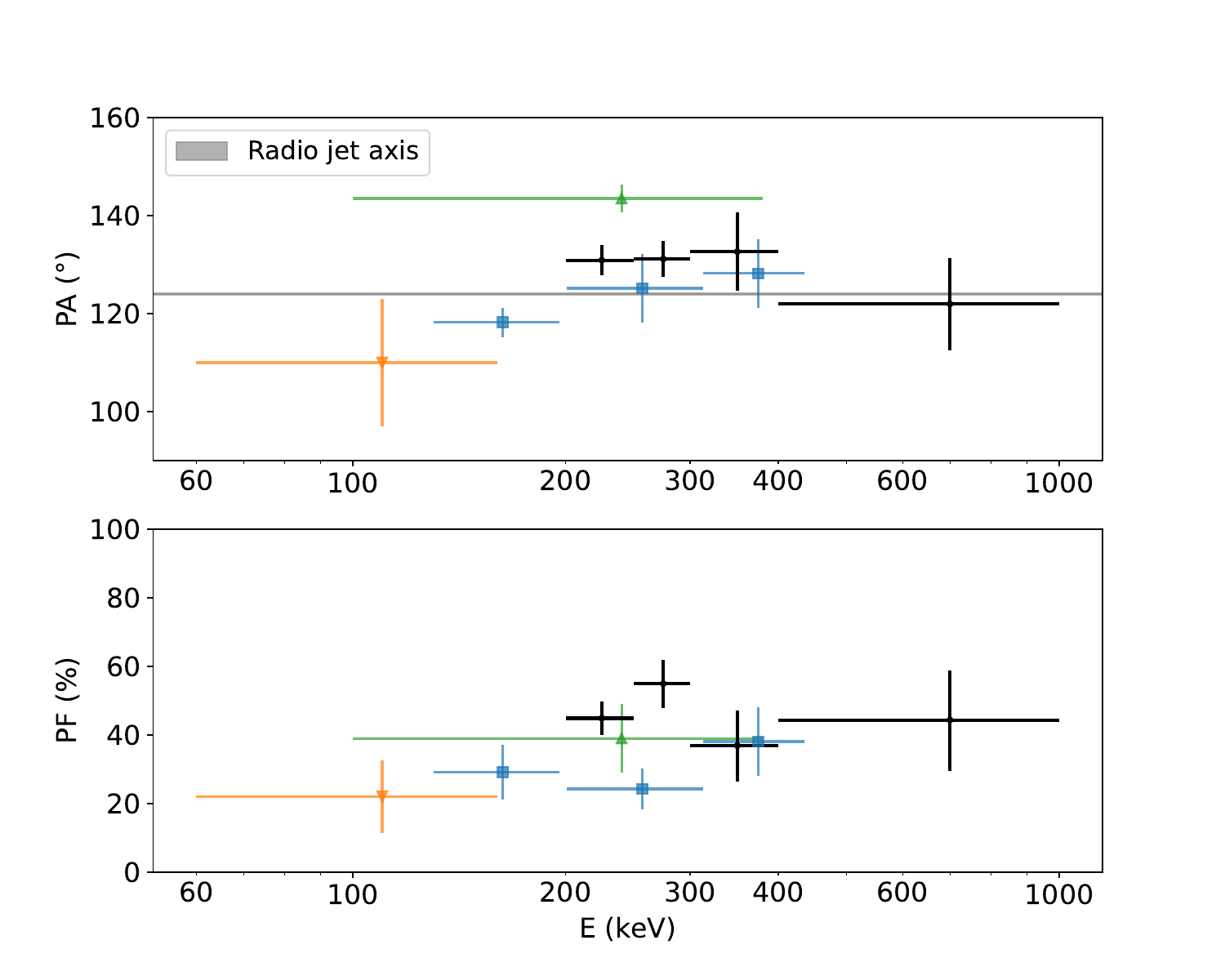}
    \caption{Evolution of the Crab Nebula polarization with energy for the entire dataset (2003--2024) in black. The other missions showed are INTEGRAL/SPI (blue), AstroSat (green) and Hitomi (orange).}
    \label{fig:crab_polarspec_2003_2024}
\end{figure}

The combination of the entire dataset is shown in Fig. \ref{fig:crab_polarspec_2003_2024} and Table \ref{tab:crab_specpolar_all}. The polarization is similar to Period 1, with the only noticeable difference being a lower PF for the 400 -- 1000\,keV band. We note that the latter is the highest energy polarization measurement ever performed for the Crab Nebula, being above 400\,keV.\\

\begin{table}[h!]
    \centering
    \caption{Crab Nebula polarization measured by the IBIS Compton mode after summing all data from 2003 to 2024.}
    \begin{tabular}{ccccc}
    E (keV) & S/N & PA (°) & PF (\%) & p-value (\%) \\
    \hhline{=====}
    & & & & \\
    200 -- 250 & 86 & 130.9 $\pm$ 3.1 & 44.9 $\pm$ 4.8 & 7e-18 \\
    250 -- 300 & 61 & 131.1 $\pm$ 3.7 & 55.0 $\pm$ 7.1 & 6e-12 \\
    300 -- 400 & 50 & 132.6 $\pm$ 8.0 & 36 $\pm$ 10 & 8e-2 \\
    400 -- 1000 & 44 & 122.0 $\pm$ 9.4 & 44 $\pm$ 15 & 4e-1 \\
    \hline
    \end{tabular}
    \label{tab:crab_specpolar_all}
\end{table}

\section{Discussion}\label{sec:discussion}

The comparison in time and energy of the Crab polarization shows that the IBIS Compton mode does not have strong systematic errors, as the data points are mainly compatible with similar experiments.\\

\subsection{On the polarization fraction}

For synchrotron emission, the maximum polarization fraction allowed can be found from the electron power-law distribution \citep{Longair_2011}. From the photon index of $\Gamma=2.24$ found in Section \ref{sec:calibration}, the maximum PF is $\Pi_{max}=77\%$ for a perfectly ordered magnetic field. All the polarization measurements we found are lower, further confirming the coherence of our results. The difference in PF is linked with the degree of order of the underlying magnetic field and can be represented by a multiplicative factor, $f=\Pi_0/\Pi_{max}$ \citep{Hughes_piston_1985}. For the combined data set, our measurement results in $f=0.65\pm0.09$ in the 200 -- 1000\,keV band, showing only moderate disordering.\\

Finally, the tension in polarization fraction between the low ($E\lesssim100$\,keV) and high ($E\gtrsim100$\,keV) energies is now firmly established, and is shown to persist in time. The change of spectral behavior around the same energy could hint at a potential explanation. In particular, it was shown that a dependence with energy of the polarization can arise for an electron distribution with a sharp spectral change and a non-uniform pitch-angle distribution \cite{Bjornsson_1985}.\\

\subsection{On the polarization angle}

While the variations of PF are easily explained by changes of the degree of order, the large discrepancies of the PA ($\sim$20°) are harder to reconcile. If the differences are intrinsic to the source, this implies that there are large changes in the magnetic field configuration, or that different parts of the nebula are emitting at different times \citep{Moran_2016}. This explanation does not hold for the measurements close in time and energy, such as AstroSat and IBIS, since in that case, the resulting polarization fraction would be lower when integrating over long periods with varying PA.\\

There could also be larger systematic errors than expected for Astrosat and/or IBIS. The PA of IBIS near MJD 57500 being compatible with POGO+ and SPI, the uncertainties over the polarization angle measured by Astrosat could be underestimated. A pulsar-phase-dependent polarization analysis would help to decipher this perplexing behavior.

\section{Conclusion}\label{sec:conclusion}

We have detailed here the complex scheme of calibration and data analysis of the INTEGRAL/IBIS Compton mode. We have shown that the Crab Nebula polarization was largely stable between 2003 and 2024, but slightly variable in energy between 200 and 1000 keV. We have compared this result with measurements made by others satellites, and shown that the polarization angle and fraction remain similar.\\

This has also enabled us to produce the first polarization measurement of the Crab Nebula and pulsar strictly above 400\,keV. This entirely new polarization constraint could potentially shed light on the emission process and geometry of the source, along with lower energy measurements. With the new calibration, we are now able to study polarization in the 200 -- 300\,keV band in detail for the first time, which will expand the number of possible sources to study. The processed product can be analyzed from a publicly available Python library, described in Appendix \ref{appendix:library}.\\

As we have seen in Section\,\ref{sec:calibration}, some issues are remaining in the estimation of the spurious flux in the lower energy band (below 350\,keV). Improving this calibration further without introducing ad hoc parameters is still an ongoing work. In particular, this requires an in-depth understanding of the detectors, their electronics and onboard data processing.\\

Obtaining reliable results through rigorous comparison and clear evaluation of systematic effects is of upmost importance in this new era of high-energy polarimetry. Results shown in this paper could in particular be used to assess the sensitivity of future high-energy polarimeters. Table \ref{tab:future_polar} shows a few future missions with polarimetric capabilities which would be able to study the Crab Nebula and pulsar polarization in details in the same energy range.\\

\begin{table}[h!]
\caption{Future missions with polarimetric capabilities, with mission status given as of October 2025.}
\label{tab:future_polar}
\centering
\begin{threeparttable}
{\renewcommand{\arraystretch}{1.3}
\begin{tabular}{cccc}
Instrument & Polarization band (keV) & Mission status\\
\hline
\midrule
COSI$^1$ & 200 -- 500 & Launch 2027 \\
COMCUBE-S$^2$ & 100 -- 460 & Phase A (ESA) \\
PHEMTO$^3$ & 50 -- 600 & ESA M8, Step 2 \\
GRINTA/HXI$^4$ & 50 -- 200 & ESA F3, Step 2 \\
\hline
\end{tabular}}
    \begin{tablenotes}
    \small
    \item $^1$\cite{cosi_2019}; $^2$\cite{comecubes_2025}; $^3$\cite{Laurent_2021_phemto}; $^4$\cite{grinta_2025}
    \end{tablenotes}
    \end{threeparttable}
\end{table}

\begin{acknowledgements}
We thank the referee for their careful reading and fruitful comments. We acknowledge partial funding from the French Space Agency (CNES). We thank E. Jourdain and J-P.Roques (IRAP) for their help using SPIDAI. We thank A. Sauvageon (CEA) and G. La Rosa (INAF) for their help on the IBIS calibration and satellite onboard selection. Based on observations with INTEGRAL, an ESA project with instruments and science data center funded by ESA member states (especially the PI countries: Denmark, France, Germany, Italy, Switzerland, Spain) and with the participation of Russia and the USA.
\end{acknowledgements}

\bibliographystyle{aa} 
\bibliography{references}

\begin{appendix}

\onecolumn

\section{Workflow chart}\label{app:workflow}

\begin{figure*}[h!]
    \centering
    \includegraphics[width=420pt]{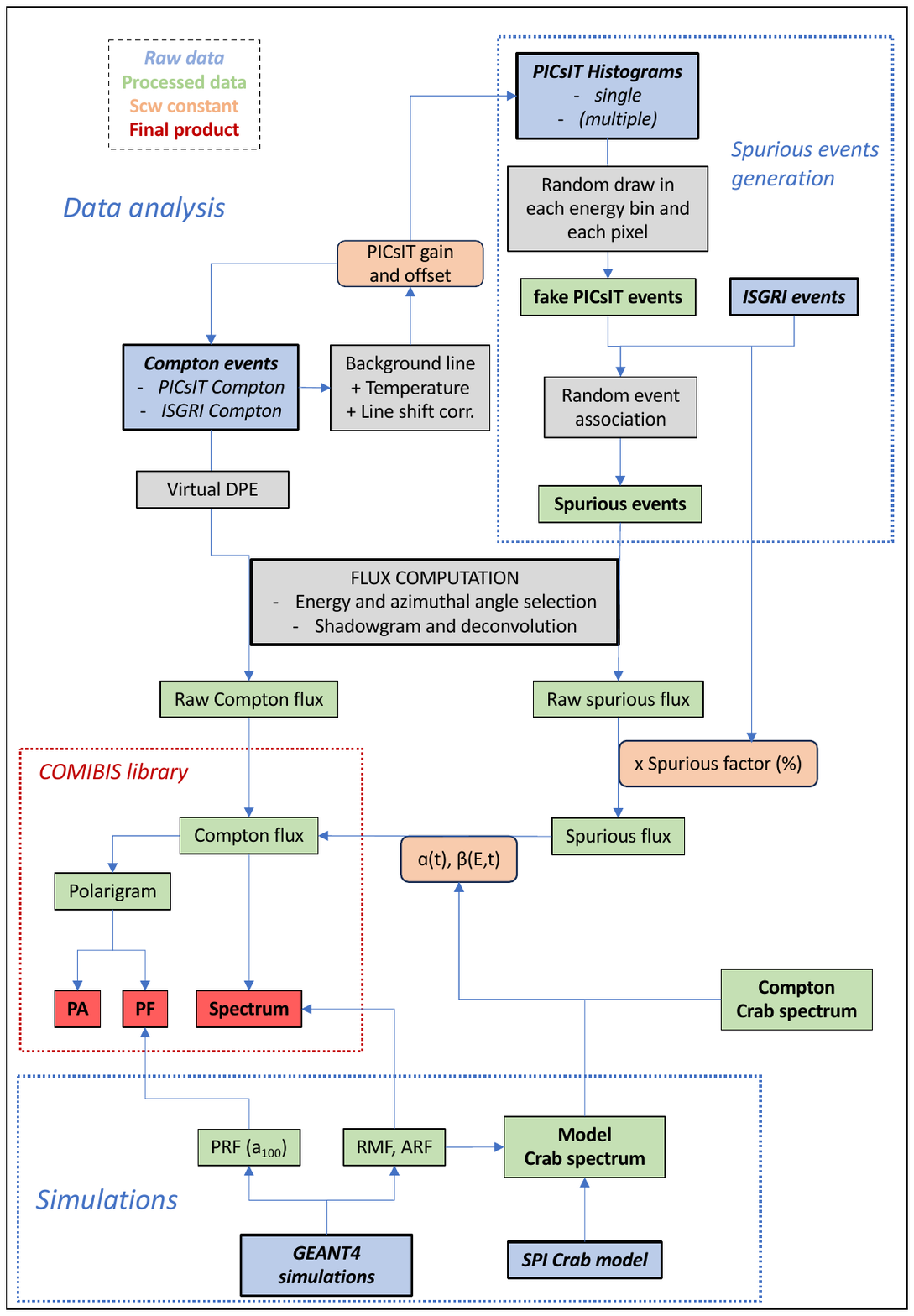}    
    \caption{Workflow chart of the IBIS Compton mode pipeline. The basic interactions with housekeeping and attitude data are omitted.}
    \label{fig:workflow_compton}
\end{figure*}

\twocolumn

\section{Marginalized posterior estimates}\label{appendix:marginal}
In Bayesian statistics, the posterior probability is the probability of the source having the polarization parameters $\vec{\theta}=(\Pi_s,\Psi_s)$ knowing the observation $\vec{X}=(\Pi_0,\Psi_0)$. This probability is computed with the Bayes theorem from the likelihood and the prior distribution. In our case, the prior is uniform for any $(\Pi_s,\Psi_s)$, and bounded between 0 and 1 for $\Pi_s$, as a source cannot be polarized above 100\%. It is possible, on the other hand, to measure $\Pi_0$ above 100\%, due to statistical fluctuations.\\

The resulting posterior probability density, $\mathcal{B}$, is almost identical to the PDF, except for $\Pi_0$ now being treated as a constant:
\begin{equation}\label{eq:2D_posterior}
        \mathcal{B}(\Pi_s,\Psi_s) = \mathcal{B}_0\frac{\Pi_0}{\pi\sigma_0^2} \exp{\left( -\frac{\Pi_0^2+\Pi_s^2-2\,\Pi_0\,\Pi_s \cos{(2(\Psi_0-\Psi_s))}}{2\,\sigma_0^2}\right)}
\end{equation}

where $\mathcal{B}_0$ is a normalization factor found by integration. This probability is marginalized on $\Pi_s$ (resp. $\Psi_s$), to compute the best value and the Highest Probability Interval (HPI) at 68\% of $\Psi_s$ (resp. $\Pi_s$). The best value is defined here as the mode of the marginalized distribution.\\

\subsection*{Marginalized posterior on polarization fraction}

For the polarization fraction, the distribution is asymmetric. With the strictly positive uniform prior, the integration on $\Psi_s$ leads to a known solution -- the Rice distribution \citep{Weisskopf_2006}. The marginalized distribution for $\Pi_s$ is:
\begin{equation}
    \mathcal{M}(\Pi_s) = \mathcal{M}_0 \frac{\Pi_0}{\sigma_0^2}\, \exp{\left(-\frac{\Pi_s^2+\Pi_0^2}{2\sigma_0^2}\right)}\,I_0{\left(\frac{\Pi_s \Pi_0}{\sigma_0^2} \right)}
\end{equation}
where $I_0$ is the zeroth order modified Bessel function of the first kind, and $\mathcal{M}_0$ is a normalizing factor found (numerically) by summing the distribution over the $\left[0,1\right]$ interval. We sampled this distribution on a fine grid ($\delta\Pi=0.001\%$) and used the \textsc{arviz} Python library to find the HPI \citep{arviz}, leading to asymmetric errors. The mode is also found at a slightly lower value than the initial $\Pi_0$, typically by a few percent.\\

\subsection*{Marginalized posterior on polarization angle}

The marginalized posterior distribution of $\Psi_s$, $\mathcal{N}$, is found by integrating Eq. \ref{eq:2D_posterior} between 0 and 1 with respect to $\Pi_s$. After a few variable changes, and making use of the fact that:
\begin{equation}
    \int_{a}^{b}e^{-t^2}dt=\frac{\sqrt{\pi}}{2}\left( \erf{(b)}-\erf{(a)} \right)
\end{equation}
where $\erf$ is the error function, we arrive at the analytical formula:
\begin{equation}
    \mathcal{N}(\Psi_{s})= \mathcal{N}_0 \exp{\left(\frac{\Pi_0^2 y^2}{2\sigma_0^2}\right)} \left( \erf{\left(\frac{1-\Pi_0 y}{\sqrt{2}\sigma_0}\right)}+\erf{\left(\frac{\Pi_0 y}{\sqrt{2}\sigma_0}\right)} \right)
\end{equation}
where $y=\cos{(2(\Psi_0-\Psi_s)})$, and $\mathcal{N}_0$ is a normalization factor found by summing the distribution on the $\left[-\pi/2,\pi/2\right]$ interval. The resulting distribution being symmetric, the errors are simply found by equating a cumulative sum above the mode to $34\%$ (or half of any desired probability). This function being periodic, it can lead to convergence issues if $\Psi_0$ is close to a boundary. To avoid this problem, we set $\Psi_0=0$ when computing the error, since the distribution only depends on the difference between the two angles.\\

\section{Confidence contours}\label{appendix:contours}

The confidence contours on the 2D plane ($\Pi_s, \Psi_s$), are first computed using the negative log-likelihood (NLL):
\begin{equation}\label{eq:polar_likelihood}
    -\ln{\mathcal{L}} = \frac{\Pi_0^2+\Pi_s^2-2\,\Pi_0\,\Pi_s \cos{(2(\Psi_0-\Psi_s))}}{2\,\sigma_0^2}+K
\end{equation}
where the constant $K=-\ln{\left(\frac{\Pi_0}{\pi \sigma_0^2}\right)}$ is the minimum value of the NLL, allowing us to define $\Delta\mathrm{NLL}$ as:
\begin{equation}
    \Delta\mathrm{NLL}=\frac{\Pi_0^2+\Pi_s^2-2\,\Pi_0\,\Pi_s \cos{(2(\Psi_0-\Psi_s))}}{2\,\sigma_0^2}
\end{equation}
whose minimum is zero, reached for ($\Pi_s=\Pi_0$, $\Psi_s=\Psi_0$). The contour at a desired level $p_c$ is found with the $\chi^2$ function with 2 degrees of freedom (dof) using Wilk's theorem approximation:
\begin{equation}
    \Delta\mathrm{NLL}=\frac{1}{2}\,\chi^2(p_c;2\,dof)
\end{equation}

Examples for $p_c=68$ and $99$\% are shown in Fig. \ref{fig:nll_cn_all_200_400_polar}.\\

\section{Crab Nebula spectral parameters}

\begin{table}[h!]
    \caption{Evolution of the Crab Nebula spectral parameters from the Compton mode fitting, after $\beta$ correction.}
    \label{tab:crab_spec_param}
    \centering
    \begin{tabular}{cccc}
    Date & $F$ (350 -- 1000 keV) & $\Gamma$ & Exposure \\
     & 10$^{-9}$erg.s$^{-1}$.cm$^{-2}$ &  &  Ms \\
    \hline
    \midrule
    2003 -- 2004 & 6.41 $\pm$ 0.44 & 1.65 $\pm$ 0.19 & 0.26 \\
    2005 -- 2006 & 6.57 $\pm$ 0.77 & 2.35 $\pm$ 0.37 & 0.31 \\
    2007 -- 2008 & 6.45 $\pm$ 0.69 & 2.73 $\pm$ 0.39 & 0.65 \\
    2009 -- 2010 & 6.60 $\pm$ 0.37 & 2.32 $\pm$ 0.19 & 0.59 \\
    2011 -- 2012 & 6.61 $\pm$ 0.36 & 2.45 $\pm$ 0.19 & 0.64 \\
    2013 -- 2014 & 6.61 $\pm$ 0.24 & 2.25 $\pm$ 0.12 & 1.02 \\
    2015 -- 2016 & 6.65 $\pm$ 0.22 & 2.32 $\pm$ 0.11 & 1.62 \\
    2017 -- 2018 & 6.59 $\pm$ 0.48 & 2.29 $\pm$ 0.24 & 1.41 \\
    2019 -- 2020 & 6.55 $\pm$ 0.51 & 2.19 $\pm$ 0.25 & 1.47 \\
    2021 -- 2022 & 6.52 $\pm$ 0.49 & 2.64 $\pm$ 0.26 & 1.55 \\
    2023 -- 2024 & 6.56 $\pm$ 0.36 & 2.49 $\pm$ 0.18 & 1.41 \\
    \hline
    \end{tabular}

\end{table}

\section{The COMIBIS Python library}\label{appendix:library}

The public python library \textsc{comibis} (contraction of COMpton mode of IBIS) was developed along with notebook examples to facilitate the analysis of raw fluxes. It is available on GitHub\footnote{\url{https://github.com/tristanbouchet/IBIS-Compton-mode-polarization}}. Users can select INTEGRAL science windows with criteria on dates, satellite revolutions, maximum off-axis angle and energy bands, then create many polarigrams, combine them and show the resulting polarization parameters along with statistical errors. Another module of the library is dedicated to spectral analysis. Similarly to polarization analysis, science windows can be selected and combined to produce spectra between 350\,keV and 3\,MeV. Those spectra can either be saved as fits, or directly fitted internally with a few basic models.

\end{appendix}

\end{document}